\documentclass[aps,prc,amsfonts,onecolumn,superscriptaddress,showpacs,showkeys,nofootinbib]{revtex4}

\usepackage{color}
\usepackage{graphics,epsfig}
\usepackage{verbatim}

\pdfoutput = 1
\usepackage{amsmath}
\usepackage{amssymb}
\usepackage{amstext}
\usepackage{mathrsfs}
\usepackage{latexsym}
\usepackage{float}
\usepackage{subfigure}
\usepackage{wasysym}
\usepackage{fancyhdr}
\usepackage{lineno}

\newcommand {\be} {\begin{equation}}
\newcommand {\ba} {\begin{eqnarray}}
\newcommand {\ee} {\end{equation}}
\newcommand {\ea} {\end{eqnarray}}

\def\gOneP{$g_1^p$ }
\def\gOneD{$g_1^d$ }

\def\dgx{$\Delta G(x)$}

\def\prc#1{Phys.\ Rev.\ C {\bf #1}}

\begin{document}
\hyphenation{RCSLACPOL}


\newcommand*{\LOSALAMOS}{Los Alamos National Laboratory, Los Alamos, NM 87545}
\affiliation{\LOSALAMOS}

\newcommand*{\ANL}{Argonne National Laboratory, Argonne, Illinois 60439}
\newcommand*{\ANLindex}{1}
\affiliation{\ANL}
\newcommand*{\CSUDH}{California State University, Dominguez Hills, Carson, CA 90747}
\newcommand*{\CSUDHindex}{2}
\affiliation{\CSUDH}
\newcommand*{\CMU}{Carnegie Mellon University, Pittsburgh, Pennsylvania 15213}
\newcommand*{\CMUindex}{3}
\affiliation{\CMU}
\newcommand*{\CUA}{Catholic University of America, Washington, D.C. 20064}
\newcommand*{\CUAindex}{4}
\affiliation{\CUA}
\newcommand*{\SACLAY}{CEA, Centre de Saclay, Irfu/Service de Physique Nucl\'eaire, 91191 Gif-sur-Yvette, France}
\newcommand*{\SACLAYindex}{5}
\affiliation{\SACLAY}
\newcommand*{\CNU}{Christopher Newport University, Newport News, Virginia 23606}
\newcommand*{\CNUindex}{6}
\affiliation{\CNU}
\newcommand*{\UCONN}{University of Connecticut, Storrs, Connecticut 06269}
\newcommand*{\UCONNindex}{7}
\affiliation{\UCONN}
\newcommand*{\FU}{Fairfield University, Fairfield CT 06824}
\newcommand*{\FUindex}{8}
\affiliation{\FU}
\newcommand*{\FIU}{Florida International University, Miami, Florida 33199}
\newcommand*{\FIUindex}{9}
\affiliation{\FIU}
\newcommand*{\FSU}{Florida State University, Tallahassee, Florida 32306}
\newcommand*{\FSUindex}{10}
\affiliation{\FSU}
\newcommand*{\GWUI}{The George Washington University, Washington, DC 20052}
\newcommand*{\GWUIindex}{11}
\affiliation{\GWUI}
\newcommand*{\ISU}{Idaho State University, Pocatello, Idaho 83209}
\newcommand*{\ISUindex}{12}
\affiliation{\ISU}
\newcommand*{\INFNFE}{INFN, Sezione di Ferrara, 44100 Ferrara, Italy}
\newcommand*{\INFNFEindex}{13}
\affiliation{\INFNFE}
\newcommand*{\INFNFR}{INFN, Laboratori Nazionali di Frascati, 00044 Frascati, Italy}
\newcommand*{\INFNFRindex}{14}
\affiliation{\INFNFR}
\newcommand*{\INFNGE}{INFN, Sezione di Genova, 16146 Genova, Italy}
\newcommand*{\INFNGEindex}{15}
\affiliation{\INFNGE}
\newcommand*{\INFNRO}{INFN, Sezione di Roma Tor Vergata, 00133 Rome, Italy}
\newcommand*{\INFNROindex}{16}
\affiliation{\INFNRO}
\newcommand*{\ORSAY}{Institut de Physique Nucl\'eaire ORSAY, Orsay, France}
\newcommand*{\ORSAYindex}{17}
\affiliation{\ORSAY}
\newcommand*{\ITEP}{Institute of Theoretical and Experimental Physics, Moscow, 117259, Russia}
\newcommand*{\ITEPindex}{18}
\affiliation{\ITEP}
\newcommand*{\JMU}{James Madison University, Harrisonburg, Virginia 22807}
\newcommand*{\JMUindex}{19}
\affiliation{\JMU}
\newcommand*{\KNU}{Kyungpook National University, Daegu 702-701, Republic of Korea}
\newcommand*{\KNUindex}{20}
\affiliation{\KNU}
\newcommand*{\LPSC}{LPSC, Universit\'e Grenoble-Alpes, CNRS/IN2P3, INPG, Grenoble, France}
\newcommand*{\LPSCindex}{21}
\affiliation{\LPSC}
\newcommand*{\UNH}{University of New Hampshire, Durham, New Hampshire 03824-3568}
\newcommand*{\UNHindex}{22}
\affiliation{\UNH}
\newcommand*{\NSU}{Norfolk State University, Norfolk, Virginia 23504}
\newcommand*{\NSUindex}{23}
\affiliation{\NSU}
\newcommand*{\OHIOU}{Ohio University, Athens, Ohio  45701}
\newcommand*{\OHIOUindex}{24}
\affiliation{\OHIOU}
\newcommand*{\ODU}{Old Dominion University, Norfolk, Virginia 23529}
\newcommand*{\ODUindex}{25}
\affiliation{\ODU}
\newcommand*{\RPI}{Rensselaer Polytechnic Institute, Troy, New York 12180-3590}
\newcommand*{\RPIindex}{26}
\affiliation{\RPI}
\newcommand*{\URICH}{University of Richmond, Richmond, Virginia 23173}
\newcommand*{\URICHindex}{27}
\affiliation{\URICH}
\newcommand*{\ROMAII}{Universita' di Roma Tor Vergata, 00133 Rome Italy}
\newcommand*{\ROMAIIindex}{28}
\affiliation{\ROMAII}
\newcommand*{\MSU}{Skobeltsyn Institute of Nuclear Physics, Lomonosov Moscow State University, 119234 Moscow, Russia}
\newcommand*{\MSUindex}{29}
\affiliation{\MSU}
\newcommand*{\SCAROLINA}{University of South Carolina, Columbia, South Carolina 29208}
\newcommand*{\SCAROLINAindex}{30}
\affiliation{\SCAROLINA}
\newcommand*{\JLAB}{Thomas Jefferson National Accelerator Facility, Newport News, Virginia 23606}
\newcommand*{\JLABindex}{31}
\affiliation{\JLAB}
\newcommand*{\UTFSM}{Universidad T\'{e}cnica Federico Santa Mar\'{i}a, Casilla 110-V Valpara\'{i}so, Chile}
\newcommand*{\UTFSMindex}{32}
\affiliation{\UTFSM}
\newcommand*{\EDINBURGH}{Edinburgh University, Edinburgh EH9 3JZ, United Kingdom}
\newcommand*{\EDINBURGHindex}{33}
\affiliation{\EDINBURGH}
\newcommand*{\GLASGOW}{University of Glasgow, Glasgow G12 8QQ, United Kingdom}
\newcommand*{\GLASGOWindex}{34}
\affiliation{\GLASGOW}
\newcommand*{\VIRGINIA}{University of Virginia, Charlottesville, Virginia 22901}
\newcommand*{\VIRGINIAindex}{35}
\affiliation{\VIRGINIA}
\newcommand*{\WM}{College of William and Mary, Williamsburg, Virginia 23187-8795}
\newcommand*{\WMindex}{36}
\affiliation{\WM}
\newcommand*{\YEREVAN}{Yerevan Physics Institute, 375036 Yerevan, Armenia}
\newcommand*{\YEREVANindex}{37}
\affiliation{\YEREVAN}

\newcommand*{\NOWUTFSM}{Universidad T\'{e}cnica Federico Santa Mar\'{i}a, Casilla 110-V Valpara\'{i}so, Chile}
\newcommand*{\NOWODU}{Old Dominion University, Norfolk, Virginia 23529}
\newcommand*{\NOWCNU}{Christopher Newport University, Newport News, Virginia 23606}
\newcommand*{\NOWUK}{University of Kentucky, Lexington, KY 40506}
\newcommand*{\NOWLANL}{Los Alamos National Laboratory, Los Alamos, NM 87544 USA}
\newcommand*{\NOWJLAB}{Thomas Jefferson National Accelerator Facility, Newport News, Virginia 23606}
\newcommand*{\NOWGLASGOW}{University of Glasgow, Glasgow G12 8QQ, United Kingdom}
\newcommand*{\NOWORSAY}{Institut de Physique Nucl\'eaire ORSAY, Orsay, France}
\newcommand*{\NOWROMAII}{Universita' di Roma Tor Vergata, 00133 Rome Italy}


\title{Precision measurements of $g_1$ of the proton and 
the deuteron with 6 GeV electrons}

\author{Y.~Prok}
     \email{yprok@jlab.org}
     \thanks{Corresponding author.}
\affiliation{\ODU}
\affiliation{\JLAB}    
\author{P.~Bosted}
\affiliation{\WM}
\affiliation{\JLAB}
\author{N.~Kvaltine}
\affiliation{\VIRGINIA}

\author {K.P. ~Adhikari} 
\affiliation{\ODU}
\author {D.~Adikaram} 
\affiliation{\ODU}
\author{M.~Aghasyan}
\affiliation{\INFNFE}
\author {M.J.~Amaryan} 
\affiliation{\ODU}
\author {M.D.~Anderson} 
\affiliation{\GLASGOW}
\author {S. ~Anefalos~Pereira} 
\affiliation{\INFNFR}
\author {H.~Avakian}
\affiliation{\JLAB}
% ...

\collaboration{The CLAS Collaboration}
\noaffiliation

\date{\today}

\begin{abstract}
The inclusive polarized structure functions of the proton
and deuteron, \gOneP and \gOneD, were measured with high
statistical precision 
using polarized 6 GeV electrons incident on a 
polarized ammonia target in Hall B at Jefferson Laboratory.
Electrons scattered at lab angles between 18 and 45 degrees
were detected using the CEBAF Large Acceptance Spectrometer (CLAS). 
For the usual DIS kinematics, $Q^2>1$ GeV$^2$ and the final-state invariant mass $W>2$ GeV, 
the ratio of polarized to unpolarized structure functions $g_1/F_1$ is found to
be nearly independent of $Q^2$ at fixed $x$. Significant resonant
structure is apparent at values of $W$ up to 2.3 GeV.
In the framework of perturbative QCD, the high-$W$
results can be used to better constrain the polarization
of quarks and gluons in the nucleon, 
as well as high-twist contributions. 

\end{abstract}


\maketitle

\section{INTRODUCTION}\label{s1}
A full understanding  of the spin structure of the nucleon, 
including the determination of the polarized gluon density \dgx\ in 
particular, is a major goal of hadronic physics.
Deep-inelastic lepton scattering from nucleons has proved over
the years to be the cleanest tool to study the short-distance 
structure of the nucleon. The pioneering experiments at SLAC,
followed by several generations of experiments at FermiLab,
SLAC, CERN, DESY, and elsewhere, have made great strides
in determining the spin-averaged parton densities of the
quarks and gluons in the proton and neutron. Starting in 
the 1970's, experiments using polarized targets have been
making steady progress in determining the spin-dependent
longitudinal
parton densities, although over a more restricted range
in momentum fraction $x$ and momentum transfer squared
$Q^2$ due to the lower luminosity available with polarized
targets. See Ref.~\cite{reviewarticle} for a recent review.
Initial studies from SLAC and CERN, borne out
with increased precision by subsequent experiments at
SLAC-\cite{E143,E155}, CERN-\cite{COMPASS}, 
DESY-\cite{HERMES}, and JLab~\cite{reviewarticle,Eg1b}, 
showed that the up and down quarks sum to only a small fraction of
total spin of the nucleon in the framework of 
perturbative QCD. 
This implied that the net contribution
of polarized gluons, strange quarks, and parton angular
momentum must be substantial. A large effort is ongoing
to study these contributions.

For gluon spin, two approaches are being
followed. The first is to try to isolate specific
processes in which a polarized gluon is involved at
leading order, for example, photon-gluon fusion leading
to a pair of charmed quarks (COMPASS)~\cite{COMPASS}, or quark-gluon
scattering leading to a high energy photon (RHIC-spin) ~\cite{RHIC}.
The interpretation of these interactions is complicated
due both to background events (other tree-level processes that
can lead to the same final state) and higher
order QCD corrections~\cite{reviewarticle}.

The second approach examines the $Q^2$ dependence of the spin structure
function $g_1$.
Perturbative QCD allows a simple 
expression of $g_1$ in terms of the
quark, antiquark, and gluon distributions 
$\Delta q$, $\Delta\overline{q}$ and
$\Delta G$, which evolve in leading order according to the
DGLAP equations~\cite{DGLAP}: 

\ba
  g_1(x,Q^2)_{_\mathrm{pQCD}} = \frac{1}{2} \sum e_q^2  \left[
\frac{}{}(\Delta q+\Delta\overline{q})\: \otimes \: \right. 
    \left.   \left (1+\frac{{\alpha}_s(Q^2)}{2 \pi}\delta C_q \right)
+\frac{{\alpha}_s(Q^2)}{2 \pi} \Delta G \: \otimes 
\frac{\delta C_G}{N_f}  \right],  
  \label{eq:pQCD}
\ea
where ${\alpha}_s$ is the strong coupling factor, ${\delta}C_q$ and ${\delta}C_G$ are the Wilson 
coefficients, $N_f$ is the number of quark flavors, and $\otimes$ indicates a convolution integral.

In the spin-averaged case, pQCD evolution  
is the benchmark approach
to which reaction-specific determinations of the gluon
density, $G(x)$, are compared. This is possible due to 
the high precision of measurements of the spin-averaged
structure function $F_2$ over many decades in both
$x$ and $Q^2$ (needed because the logarithmic evolution
due to gluon radiation is very slow).
In the polarized case, the kinematic range of present
precision data is
considerably more limited. Nonetheless, the data are
of sufficient quality to obtain a very good description
of the valence up and down quark densities, and rough
indications of the gluon and sea quark 
densities~\cite{reviewarticle}.

In practice, fits to data should include the effects of both
kinematic and dynamic higher twist, especially in the pre-asymptotic
regime of this experiment. We can decompose $g_1(x,Q^2)$ into a leading twist and a
higher twist component~\cite{lss-2007} :
\be
g_1(x,Q^2) = g_1(x,Q^2)_{LT}+g_1(x,Q^2)_{HT}.
\ee
The leading twist term can be expressed as a sum of the 
well-known pQCD term 
$g_1(x,Q^2)_{_\mathrm{pQCD}}$~\cite{reviewarticle},
the calculable kinematic target mass corrections, and higher order terms:
\be
g_1(x,Q^2)_{LT} = g_1(x,Q^2)_{_\mathrm{pQCD}}+h^{TMC}(x,Q^2)/Q^2 + \mathcal O(1/Q^4).
\ee
The dynamic higher twist contribution represents the parton-parton correlations in the target nucleon
and is given by:
\be
g_1(x,Q^2)_{HT} = h(x,Q^2)/Q^2+ \mathcal O(1/Q^4).
\ee
The dynamic higher twist corrections are non-perturbative in nature 
and cannot presently be calculated in a model-independent way. 
However, they can be extracted directly from the experimental data, 
which has been done recently with CLAS and COMPASS data by the LSS group~\cite{lss-2007}.

The goal of the present experiment is to provide much
higher statistical precision in measurements of \gOneP 
and \gOneD  than was previously available
in the kinematic range accessible with 6 GeV
electrons at Jefferson Lab. 
This precision is needed
to distinguish between power-law higher twist 
and logarithmic gluon radiation in the polarized  parton densities in the nucleon. 
Much of the present data also lie in the pre-asymptotic region, and exhibit significant 
resonance structure, which needs to be taken into account by QCD fits including the higher twist terms. 
These data will be particularly valuable 
when combined with planned data with 11 GeV electrons.

\section{THE EXPERIMENT}\label{s3}
\subsection{Experimental Context}
This experiment has been carried out at Jefferson Lab, 
using longitudinally polarized electrons incident on
longitudinally polarized protons and deuterons in 
solid ammonia targets,
and the CEBAF Large Acceptance Spectrometer (CLAS)~\cite{CLAS}. 
Data from several previous
experiments with solid polarized targets have been 
analyzed and published \cite{eg1a,Eg1b},
and some are still in the analysis stage.

The present experiment used 6 GeV electrons and relatively 
larger scattering angles to focus
on the Deep Inelastic region (DIS). The main focus of this experiment was
on semi-inclusive DIS (SIDIS) and deeply virtual Compton
scattering (DVCS), both of which required the detection
of photons at small angles. For this reason, an
Inner Calorimeter (IC) was installed, which blocked
scattered electrons below about 18 degrees.
The IC is not used in the present inclusive analysis.

The present analysis closely follows those of its 
predecessors~\cite{eg1a,Eg1b}.
Small differences arise from the use of $^{14}$NH$_3$
instead of $^{15}$NH$_3$, a 50\% longer target cell, 
the inclusion of the IC detector into CLAS, and an improved treatment of the effects of the polarized
target magnetic field.

\subsection{Experimental Overview}

The experiment, named  ``eg1-dvcs'', used 
6 GeV longitudinally polarized electrons 
from CEBAF at Jefferson Lab impinging
on a longitudinally polarized 
solid ammonia target of 0.025 radiation length, immersed in liquid helium 
~\cite{Keith}. Inclusive scattered electrons
were detected in CLAS~\cite{CLAS}.
The typical beam current was 7 nA, which, when integrated over the 6 months
of data taking, resulted in approximately 
$2\times 10^{17}$ electrons traversing the target.
The beam polarization, as periodically measured using
M\o ller scattering in an upstream polarimeter, averaged
85\% for the first three quarters of the experiment. A lower
polarization of about 75\% was delivered during the
remaining time. 

About 70\% of the running time was on polarized
protons (NH$_3$ target), 20\% on polarized deuterons
(ND$_3$ target), 10\% on a reference unpolarized
carbon target, and 1\% on an empty cell. The 1.5-cm-diameter cells typically contained
1 g/cm$^2$ of material immersed in a 2-cm-long
liquid helium bath.  The target insert housed 2 such cells, which were called the "top", and 
the "bottom" cells. In order to minimize loss of polarization from beam heating, 
the sub-millimeter-diameter beam was uniformly rastered over
the front face of the target with a period of a few seconds.

The
beam position, averaged over a few minutes or longer,
was kept stable at the 0.1 mm level, using feedback from
a set of beam position monitors. 
A split solenoid superconducting magnet provided a highly
uniform 5-T magnetic field near the target, which
effectively extended about 20 cm up- and downstream 
from the target center.

\begin{figure}[hbt]

\centerline{\includegraphics[width=2.4in]{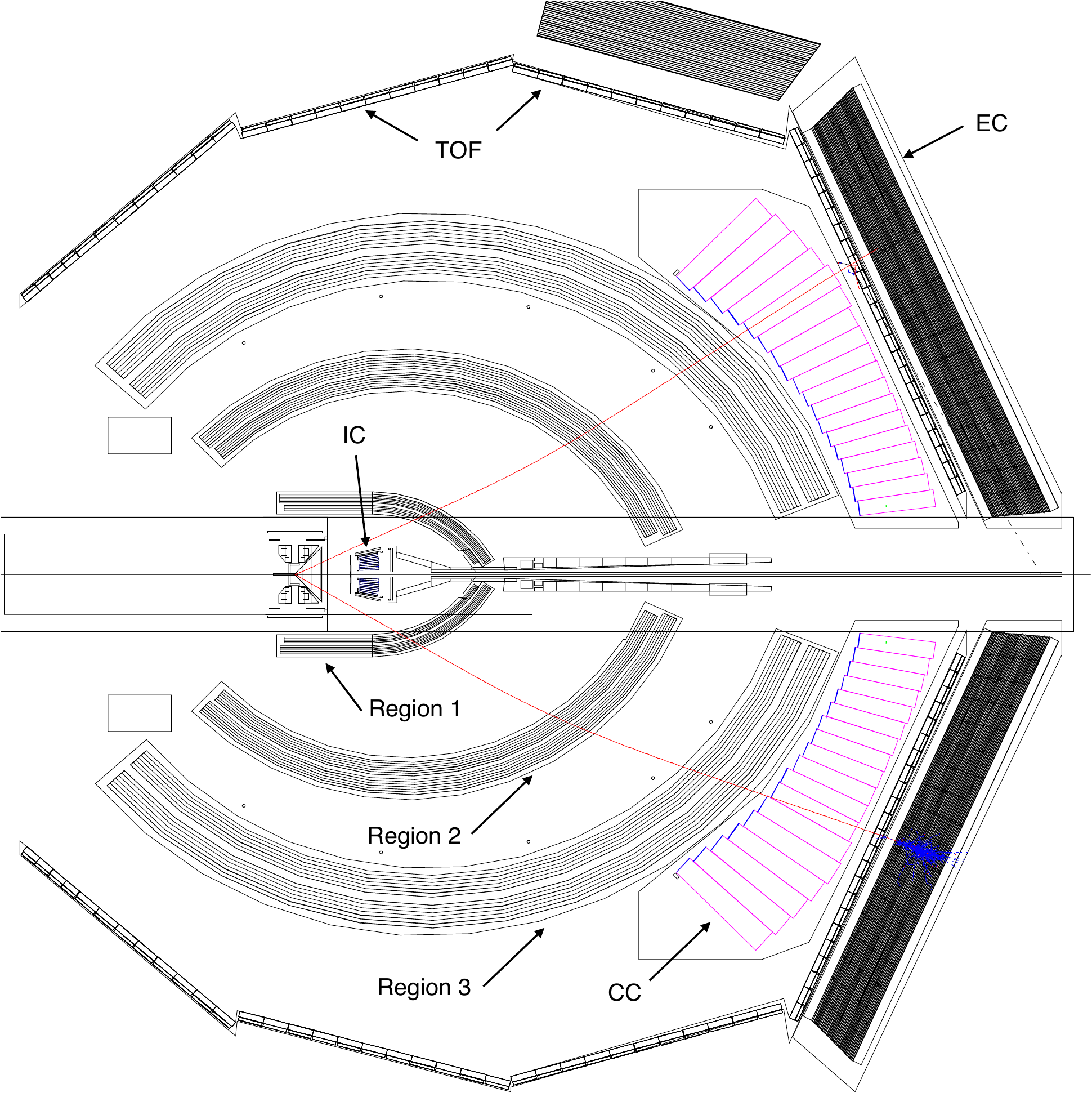}}

\caption{ (color online) A schematic of an event in CLAS, with 2 charged particles
passing through 3 regions of drift chambers (DC), 
the time of flight (TOF, also known as SC) 
the Cherenkov Counters (CC),
and the electromagnetic calorimeter (EC).
}
\label{fig:CLAS}
\end{figure}

Scattered electrons were detected in the 
CLAS detector ~\cite{CLAS} in Hall B, over the polar angle 
range of about 18 to 48 degrees. A typical event in CLAS is
shown in Figure ~\ref{fig:CLAS}.

The CLAS, comprised of six azimuthally symmetric detector
arrays,  uses a
toroidal field to bend charged particles. Particle momenta and scattering angles
were measured with a drift chamber (DC) tracking system with a
relative accuracy of 0.3\% to 2\% in momentum, and
about 3 mr in angle.
Electrons were separated from a significantly 
larger flux of charged pions using segmented
gas Cherenkov detectors (pion threshold 2.6 GeV)
and an electromagnetic calorimeter. A hardware
trigger system made the first selection, rejecting
about 90\% of pions while keeping close to 99\%
of electrons, in order to not overwhelm the 
data acquisition system. The hardware Cherenkov
and calorimeter thresholds were adjusted to give
a trigger rate of about 3000 Hz, and a ``dead time''
in which no trigger could be accepted of about 10\%.
An additional unbiased trigger was prescaled by
a large factor in order to measure the efficiency
of the main electron trigger.

The data taking was divided into three parts: Part A
in early 2009 used NH$_3$ as the target, centered 
at 58 cm upstream of CLAS center ($z_0=-58$ cm); 
Part B (mid 2009) also
used NH$_3$, this time at $z_0=-68$ cm; and Part C (September
2009) mostly used ND$_3$ as the 
target, again with  $z_0=-68$ cm.
Each part had slightly different primary beam energies 
(between 5.7 and 6.0 GeV, with several days at 4.8 GeV at the end of Part A). 
The CLAS torus polarity was set to bend electrons inwards for almost all of parts
A and B, and about two thirds of Part C.
The field strength was two thirds of the maximum available (2250 A),
resulting in better acceptance for low momentum charged particles,
but somewhat worse momentum resolution.


\section{DATA ANALYSIS}\label{s4}

For the present analysis, the only particles of
interest were scattered electrons. The spin structure
functions were determined from the difference in
rates in a particular $x$ and $Q^2$ bin for beam
and target polarizations aligned or anti-aligned. A large
background of scattering from unpolarized (or
slightly polarized) material in the target was taken
into account using a detailed model of the target composition, 
combined with a fit to world cross section data.

\subsection{Data processing}
The raw data consisted of approximately 50,000 files,
each about 2 GB in size and corresponding to a few minutes
of data taking. The files were grouped into runs
consisting of about 80 files, on average, all with the
same experimental configuration. Each file was 
processed with the standard CLAS analysis package~\cite{InclNote}.

The processed data were corrected using an improved
method of tracking particles through the target
magnetic field. The standard method in the CLAS reconstruction package swims particles backwards
to a plane perpendicular to the appropriate sector.  
A more accurate method
is to swim back to the $x$ and $y$ coordinates specified by the raster magnet.  
This method used the track coordinates
at the first layer of Drift Chambers and a fit to a large ensemble of simulated
forward-swimming particles. This ensured that
the track would intersect the beam line for a given
raster magnet setting. The net result was about a 
factor of two improved angular resolution, as verified
by the width of the beam energy, $E_0$, reconstructed from the electron
and proton scattering angles in $ep$ elastic scattering.

\subsection{Calibrations}

	\subsubsection{Standard Calibrations}
	Standard calibration procedures were done for each of 
	the subsystems in CLAS. These included: drift chamber (DC)
	alignment using straight-through tracks; DC
	timing alignments; 
	gas Cherenkov detector (CC) pulse height 
	alignment using single photo-electron peaks ; 
	time-of-flight scintillator (SC) timing corrections; and
	electromagnetic calorimeter (EC) pulse height corrections
	using cosmic rays. Timing and position resolutions
	for each of the systems, after calibration, were
	similar to those obtained during other electron
	running periods~\cite{CLAS}. Calibrations were made
	frequently enough to ensure very good stability
	for these detector systems.

	\subsubsection{Raster Corrections}
	An additional calibration specific to experiments with 
	polarized targets was done. This was to calibrate the beam
	position at the target, which depended on steering
	in the accelerator, as well as the strength of the magnetic field
	 used to raster the beam. 
  The $x$ and $y$ magnet field (in units
	of ADC counts) was recorded for each event trigger.
	By minimizing the width of the reconstructed
	target position along the beam line ($z$), the
	relation between magnet current and beam position
	was determined, as well as the beam position relative
	to the CLAS center for no raster magnetic field. The fit also 
  determined the target center along
	the beam line relative to the center of CLAS:
  -57.95 cm for Part A, and -67.95 cm for 
	parts B and C of the run.

	\subsubsection{Magnet Angle Correction}
	Another calibration specific to this experiment was
	the determination of the orientation of the target
	solenoid with respect to the beam line. The primary
	method used was to make the opening angle of 
	reconstructed electron-positron pairs (from photon
	conversions) as close to zero as possible, independent
        of azimuthal angle of the pair relative to the beam
        line axis.  The result of the study
	was an approximately 3 mr tilt, resulting in a
	significant polar deflection of charged particles,
	on top of the azimuthal rotation characteristic
	of a solenoidal field. An analysis using the over-constrained
        kinematics of exclusive $ep$ elastic scattering confirmed
        the 3 mr tilt.

\subsection{Event selection}
\begin{figure}[hc]
\hskip-1.1in
\centerline{\includegraphics[width=5.2in]{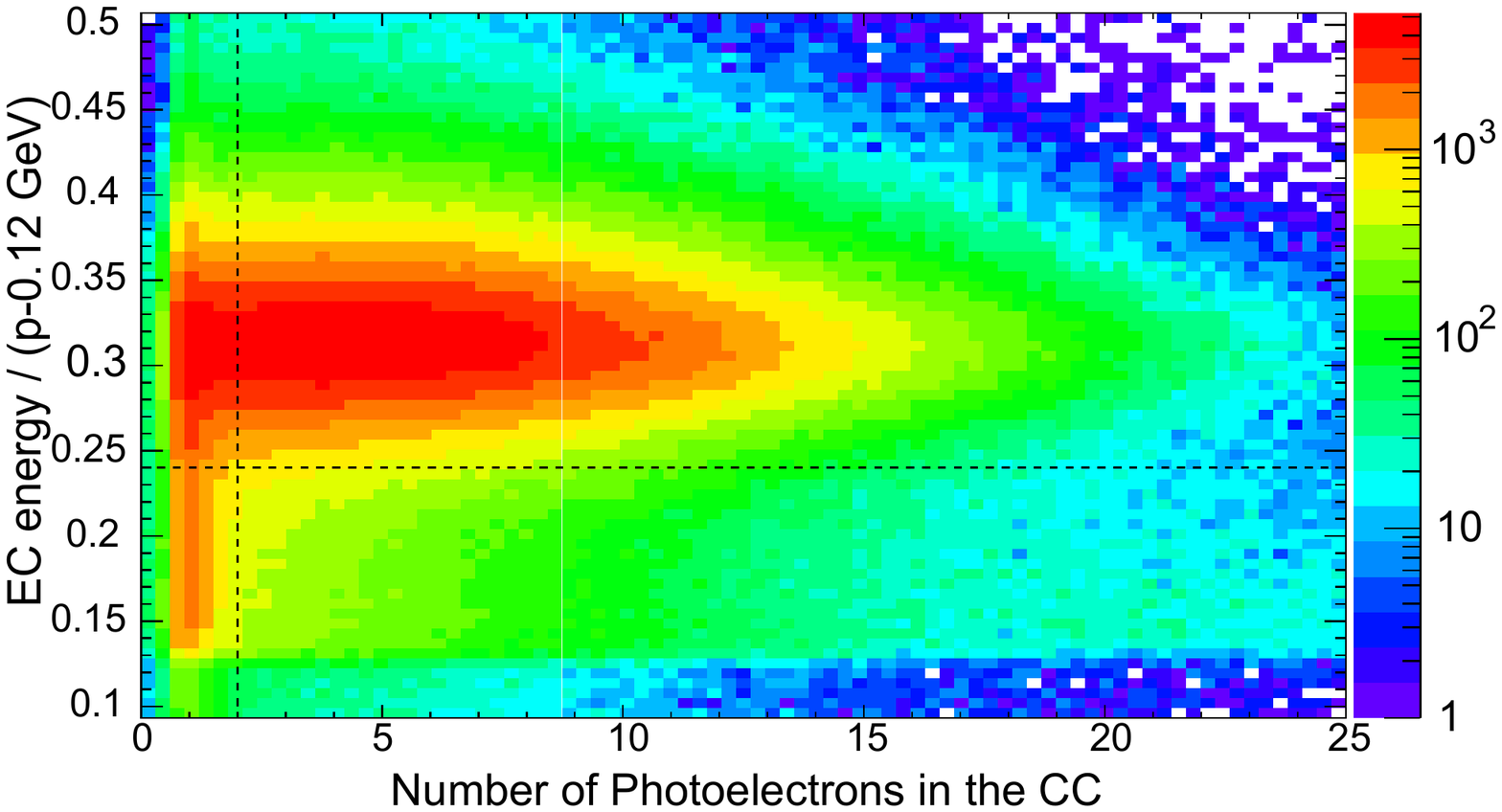}}
\vskip -1.1in
\caption{(color online) Distribution of electron candidates in the
EC energy divided by momentum $P$ (with 0.12 GeV offset) plotted vs the
number of photo-electrons in the Cherenkov detector. The dashed
lines show the cuts applied.  }
\label{fig:Eop_vs_nphe}
\end{figure}

In this experiment, the events of interest are those
with at least one well-identified electron originating
from the target. The two detectors used for particle
ID were the threshold gas Cherenkov detector
(CC, pion threshold of 2.6 GeV) and the lead-scintillator
electromagnetic calorimeter (EC), with a nominal sampling
fraction of 0.30. The first level of selection was
in the electronic trigger, which required about 1 photo-electron (p.e.)
in the CC and an energy deposition of 0.5 GeV in the EC.
In software, an electron was identified by first requiring that a time-based drift-chamber (DC) 
track have negative charge, use at
least five of the six super-layers in the tracking system,
and have a signal in each of the CC, EC, and scintillator time-of-flight
counter system (SC).
A further requirement was that the CLAS sector number for each the subsystems (DC, CC, 
SC, and EC) be matched. We then required eight additional
particle ID cuts:
\begin{itemize}
\item 
the number of photo-electrons (p.e.) in the CC must be greater than 2 p.e.
\item
The energy in the EC divided by momentum $P$ (with 0.12 GeV offset)
must be greater than $0.80 (E/P)_{peak}$, where $(E/P)_{peak}$
is the peak of the $E/(P-0.12)$ distribution (from a first
pass through the data), averaged over each run, for
each sector separately. This procedure was performed to
take out slow drifts in the EC gain, which were especially
evident in rare cases when the high voltage on the PMTs had been off
for a significant period of time. The offset of 0.12 GeV was obtained from a fit to data 
with $1.2<P<5$ GeV, and can be attributed to the energy losses in the detectors in front of the EC, 
as well as the minimum ADC signal size used in this analysis.  The distribution of EC energy
divided by momentum $P$ (with 0.12 GeV offset) plotted vs the
number of photo-electrons in the Cherenkov detector can be seen in Fig.~\ref{fig:Eop_vs_nphe}.

\item
The target vertex position along the beam line ($z$) must be within
3 cm of the polarized target center.

\item
The difference ${\delta\phi}_{DC1}$ between the electron's polar
angle at the first drift chamber layer and the azimuthal component of its 
momentum at the same location must be zero within $\pm 4^\circ$. 
Particles with higher  ${\delta\phi}_{DC1}$ are more likely to have scattered from materials 
that were not part of the target and so can have inaccurate kinematics.

\item 
The Cherenkov mirror number must be aligned with the value expected
from the track trajectory (as defined by the nearby SC paddle
number) within 2 SC paddles.
This method
uses the feature that the SC paddles and CC mirrors are
close together so that their correlation is relatively
insensitive to the track trajectories.
\item
The Cherenkov signal time must agree with the EC signal time
within $\pm$ 4 nsec.
\item 
The electron scattering angle $\theta_e$ must be less than 40 degrees for Part A, and
less than 45 degrees for parts B and C, in order to eliminate large angle particles
that could have passed through the significant amounts of the heat shield and super-insulation
in the target.
\item
The electrons' trajectories must not pass too close to the
lead shielding around the IC (or the IC support stand).
The reason for this
cut is that electrons passing through short lengths 
of the shielding could survive, but with strongly
modified angles and momenta from multiple scattering
and Bremsstrahlung, thus resulting in an incorrect
determination of their original kinematic values.
\end{itemize}

\subsection{Quality Checks}

\subsubsection{Beam Scraping}
Careful on-line monitoring ensured that very few data were
taken with the beam scraping on the relatively thick
and unpolarized target side walls. An off-line check
revealed a few files with this problem,
which were removed from further analysis.

\subsubsection{Rate Stability}
In order to avoid significant corrections to 
the measured asymmetries, and also to obtain reliable dilution
factors from the comparison of ammonia and carbon
target rates, we removed data files where an obvious
drop in efficiency occurred. 
The most common reasons for the loss of efficiency were that
parts of the DC or EC tripped off.

\subsubsection{Asymmetry Stability}
\begin{figure}[h]
\centerline{\includegraphics[width=3.6in, angle =90]{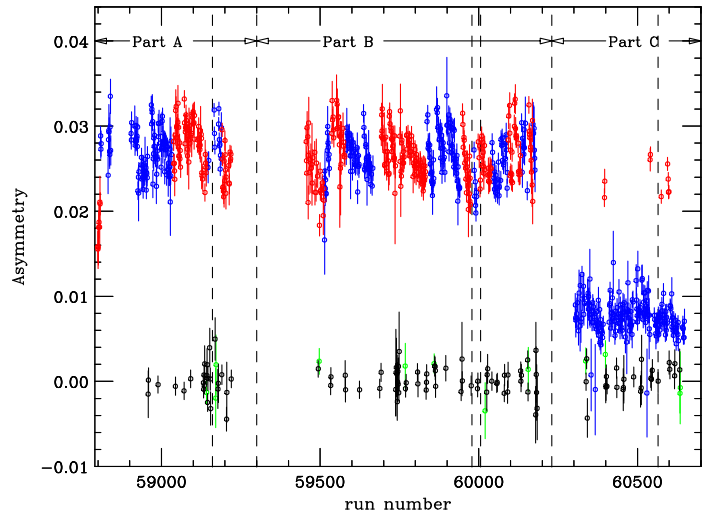}}
\caption{(color online) Average raw double-spin count asymmetries versus 
run number. The black points are for carbon, 
the green points for the empty cell,
the red points for the bottom NH$_3$ target, and the blue points for the top NH$_3$
target (Parts A and B) or the ND$_3$ target (Part C).
The vertical dashed lines correspond to changes in
beam energy or torus polarity.
}
\label{fig:asy}
\end{figure}

The study of the average inclusive electron double-spin
asymmetry proved to be a very valuable quality check.
This asymmetry is zero for carbon targets, and generally
at least 3-sigma difference from zero for polarized
target runs. The sign of the asymmetry indicated whether 
the overall beam and target polarized signs were correct
in the database.
Figure~\ref{fig:asy}
shows the final results for the double spin asymmetry
as a function of run number. In Part C, the NH$_3$
and ND$_3$ targets could also be distinguished due to
the much smaller asymmetries for ND$_3$ compared to NH$_3$.
After all corrections, 
the polarized target runs all show a positive asymmetry,
while the carbon runs are all consistent with zero.

	\subsubsection{Electron detection efficiency}
The electron detection efficiency was determined by 
comparing the rates from the
carbon target to those expected, based on fits to world 
data ~\cite{Adep}. Radiative effects and pair-symmetric
backgrounds were taken into account.  The typical efficiency was 
 $\sim$75\%, except for Sector 5, 
where it was only $\sim$40\%, due to a leak 
 in the Cherenkov detector. The efficiency was found to decrease
with increasing beam current, at a relative rate of about 1\%/nA.
This decrease is attributed to higher drift-chamber occupancy
at higher beam current, resulting in lower tracking efficiency.

\subsubsection{Beam Energy}
The beam energy was determined using measurements from the other
two experimental halls at JLab, which accurately measure the
deflection of the beam in dipole magnets. These results were
confirmed using the over-constrained kinematics of exclusive
$ep$ elastic scattering.
The results are $E_0= 5.887$, $E_0= 5.954$,
and $E_0= 5.752$ GeV for Parts A, B and C, respectively. The estimated uncertainty is 5 MeV.

\subsection{Double-spin asymmetries}
The double-spin asymmetry $A_\|$ was formed in
each two-dimensional physics bin using:
\be
A_\|^p= \frac{N^{\uparrow\Downarrow} - N^{\uparrow\Uparrow}r_c}
{N^{\uparrow\Downarrow} + N^{\uparrow\Uparrow}r_c}
\frac{c_s}
{f \hskip .05in (1 + c_1) 
\hskip .05in P_b 
\hskip .05in P_t
\hskip .05in f_{RC}  } + A_{RC},
\label{Eq:Aparrp}
\ee
for NH$_3$ and:
\be
A_\|^d= \bigg(\frac{N^{\uparrow\Downarrow} - N^{\uparrow\Uparrow}r_c}
{N^{\uparrow\Downarrow} + N^{\uparrow\Uparrow}r_c}
\frac{c_s}
{f \hskip .05in (1 + c_1)
\hskip .05in P_b 
\hskip .05in P_t} - c_2 A_\|^p \frac{\sigma_p}{\sigma_d}\bigg)
\hskip .05in \frac{1}{f_{RC}}  + A_{RC},
\label{Eq:Aparrd}
\ee

\noindent for ND$_3$, where $N^{\uparrow\Uparrow}$ ($N^{\uparrow\Downarrow}$) are the number of
counts in the parallel (anti-parallel) beam helicity
bins, $r_c$ is the ratio of incident beam charges
for the two helicities, $f$ is the bin-averaged
dilution factor, defined as the ratio of events from
polarized protons or deuterons in the NH$_3$ or ND$_3$
target to the total number of events, $c_s$ is the pair-symmetric correction,
$c_1$ accounts for polarized nitrogen
in the NH$_3$ and ND$_3$ targets, $c_2$
accounts for the polarized NH$_3$ mixed into 
the ND$_3$ target, $P_b$ is the beam polarization, $P_t$ is the target polarization, 
$A_\|^p$ is the measured value of the proton asymmetry, $A_\|^d$ is the deuteron asymmetry,
$f_{RC}$ is a
radiative dilution factor, and $A_{RC}$ is an additive radiative correction.
Each of the factors in Equations~\ref{Eq:Aparrp} 
and \ref{Eq:Aparrd} is discussed below. 
Numerical values can be found in the tables of final results in the
Appendix, and are also stored in the CLAS Physics Database \cite{clasdb}.

\subsection{Raw Asymmetry}
Data processing consisted of determining the
number of electron events passing the particle ID cuts mentioned
above, for each run, in bins of momentum $P$, 
polar scattering angle $\theta$, and helicity
(beam and target polarizations either aligned or
anti-aligned). The main reason for choosing 
bins in $(P,\theta)$ at
the data processing stage, was to allow flexibility
later on to slightly vary the beam energy $E_0$ from
the nominal value.

Counts were summed for each target and each run period:
Part A, 5.9 GeV; Part B, in-bending;
Part B, out-bending; Part C, in-bending; and
Part C, out-bending.

The bin sizes were 0.04 GeV in momentum and 0.2 degree
in $\theta$, chosen to be small enough to allow 
re-distribution into
physics bins of $(x,Q^2)$ and $(W,Q^2)$, 

\be
W =\sqrt{M^2+2M\nu-Q^2},
\label{Eq:invmass}
\ee
where $W$ is the invariant mass of the final state. 

Here, $\nu$ is the virtual photon energy, equal to the 
difference in the incident and scattered electron energies: $\nu = E_0-E^{\prime}$, 
\be
Q^2 = -q^2 = 4E_0E^{\prime} \sin^2 \frac{\theta}{2}
\ee
is the four-momentum transfer, and 
\be
x = \frac{Q^2}{2M\nu},
\ee
is the Bjorken scaling variable.
During the re-distribution process, the average values of
all relevant physics quantities were calculated.

\subsection{Correction Factor $r_c$}
The ratio of incident charges for the two beam helicity states takes into account any difference in the 
integrated incident current for one state compared to the other. Since the 
beam helicity was reversed at 30 Hz, and came in alternating helicity buckets, the 
difference came about only due to helicity-dependent beam current differences, which were kept to less 
than 1 part in 1000 on a few minute scale by a feedback system, and less than 1 part in 10000, 
averaged over a month-long time scale. Therefore, effectively $r_c$ = 1.

\subsection{Target and Beam Polarization}
For this inclusive analysis, only the 
product of target polarization ($P_t$) and beam
polarization ($P_b$) is important, but not the individual values. For the NH$_3$ targets,
the individual measurements of each quantity (NMR
for $P_t$ and M\o ller scattering for $P_b$) had relative
systematic uncertainties of over 4\%. We therefore used
$ep$ elastic scattering, for which the uncertainty on the 
measured asymmetry is of order 1\% statistical and $<2\%$
systematic. Another advantage is that the same runs
were used as for the inclusive analysis, so any
run-dependent biases would tend to cancel.

We selected electrons using similar particle ID cuts as those used for 
the inclusive analysis. We selected protons using
a $\pm 0.7$ nsec cut on the difference in predicted
and measured times between the electron and proton,
determined with the SC system. The cuts used to 
select $ep$ elastic events were:
\begin{itemize}
\item The missing energy must be less than 120 MeV
\item The missing longitudinal (transverse) momentum relative to
the incident beam direction must be less than 120 MeV (80 MeV).
\item $|W-M|<0.08$ GeV
\item The beam energy reconstructed from electron and proton
scattering angles only must agree with the nominal beam energy
to 70 MeV.
\end{itemize}
The last cut was especially powerful in reducing the
background from quasi-elastic events from nitrogen and helium to
about 3\%. 
The dilution-corrected double-spin asymmetry was then
formed for each of seven $Q^2$ bins, and $P_bP_t$ was extracted
from the ratio to the theoretically well-known 
asymmetry for $ep$ elastic scattering. The values
were then averaged over $Q^2$, which ranged from
about 2 to 7 GeV$^2$. No dependence on $Q^2$ was
observed. The results are consistent with the product
of the individual measurements of $P_b$ and $P_t$.

 \begin{table}[hbt]
 \centering
 \begin{tabular}{llc}
 Run Period & Target & $P_bP_t$  \\
\hline
  Part A (5.9 GeV)      & NH$_3$ & $   0.637\pm   0.011$ \\    
  Part B (in-bending)   & NH$_3$ & $   0.645\pm   0.007$ \\
  Part B (out-bending)  & NH$_3$ & $   0.579\pm   0.037$ \\
  Part C (in-bending)   & NH$_3$ & $   0.50\pm   0.04$ \\ 
  Part C (out-bending)  & NH$_3$ & $   0.51\pm   0.06$ \\ 
  Part C (in-bending)   & ND$_3$ & $   0.216\pm   0.010$ \\ 
  Part C (out-bending)  & ND$_3$ & $   0.236\pm   0.010$ \\ 
\hline
\end{tabular}
\caption{Average values of proton and deuteron $P_bP_t$ for all run periods. 
The results for Part C and NH$_3$ target are
for the $\sim 11\%$ of NH$_3$ contaminating the ND$_3$ target.
}
\label{tab:pbpt}
\end{table}

For the ND$_3$ target, the kinematic region where 
there is the best sensitivity to $ed$ quasi-elastic
scattering was heavily contaminated by the polarization
of the 10\% NH$_3$ in the target. On the other hand,
the direct deuteron polarization measurements using
NMR are more accurate than for the proton, thanks to
the ``double peak'' fitting method, which removes the
sensitivity to the hard-to-measure thermal equilibrium
signal~\cite{DULYA}.

We therefore used the count-weighted product
of target polarization from NMR and beam
polarization from M\o ller measurements.
We estimated the systematic uncertainty of 3\% in $P_t$ and 4\% in $P_b$. 

Using these values as a constraint, we then fit the
$ep$ coincidence data in the region of small missing
momentum to obtain $P_bP_t=0.50$ for the protons in the
ND$_3$ target.
The results for all targets  are given for each set
of running conditions in Table~\ref{tab:pbpt}.

\subsection{Dilution Factor}

\begin{figure}[hbt]
\centerline{\includegraphics[height=3.5in]{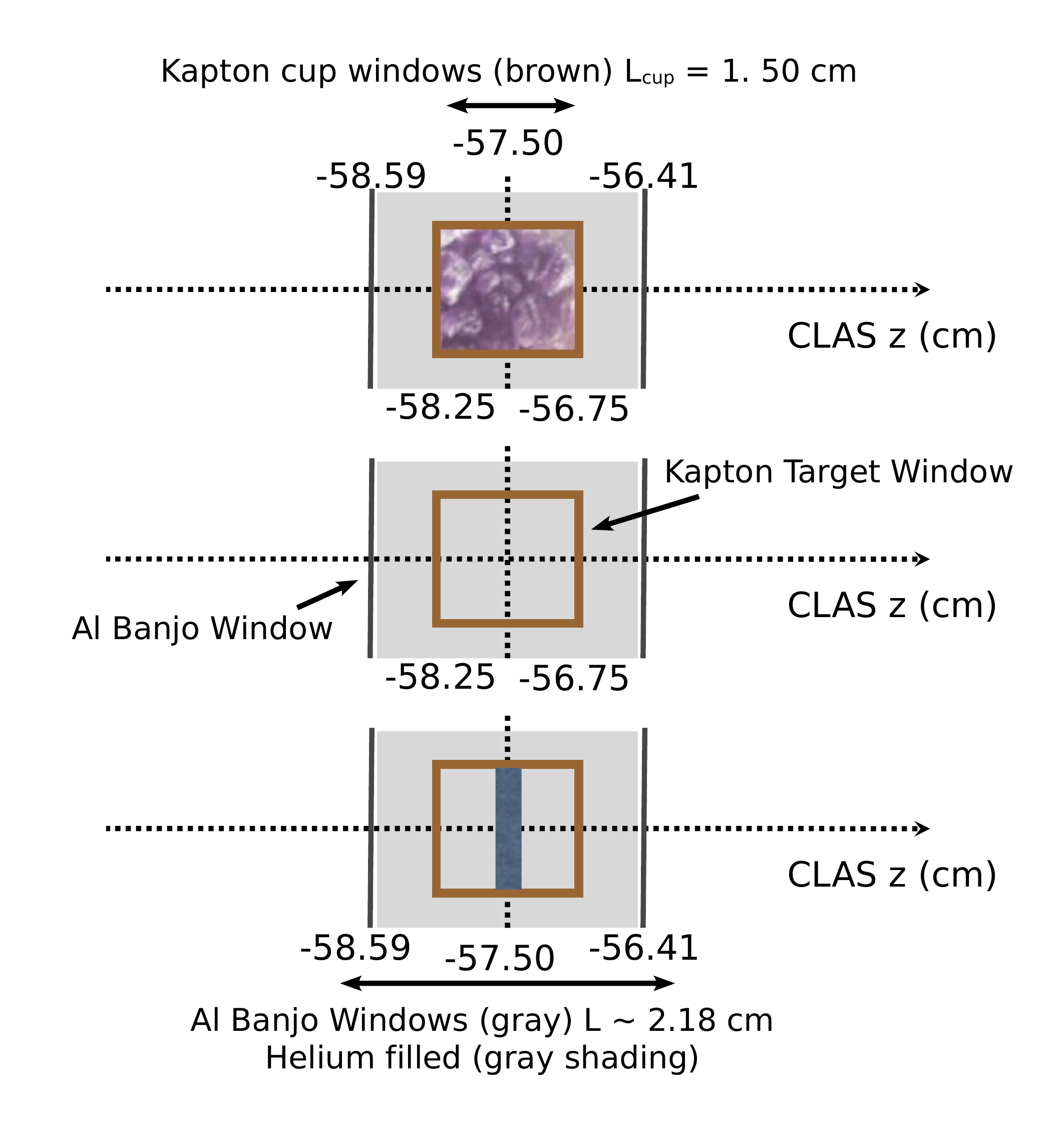}}
\caption{(color online)
Schematic diagram for eg1-dvcs showing the 
targets used: loosely packed 
ammonia beads (top), an empty target cell (middle), 
and a solid carbon target (bottom). 
Ammonia was the primary experimental target used; carbon and empty 
targets were used to calculate dilution factors and to make consistency checks.
Two ammonia cells were used during the experiment (one is not shown here).}
\label{fig:target}
\end{figure}

The dilution factor, $f$, is defined as the fraction of inclusive 
scattering events originating from polarized hydrogen or deuterium, as compared to the total number of scattering events from all materials in the target. The electron beam in our experiment 
passed through: helium (He), Kapton (K), ammonia ($^{14}$NH$_3$/$^{14}$ND$_3$), and aluminum (Al).
A schematic diagram of the three target cells is shown in Figure~\ref{fig:target}.

If we define $n_{material}$ 
as the electron scattering rate from a particular target material, we can write for the proton:
\begin{equation}
f = \frac{n_{p}}{n_{NH_3} + n_{He} + n_{Al} +n_{K}},
\end{equation}
The event rate for each material $i$ is proportional to the product of the areal density $\tilde{\rho}$ and the
inclusive DIS cross section $\sigma$: 

\begin{equation}
n_i  \propto \tilde{\rho_i} \sigma_i  \propto \rho_i l_i \sigma_i, 
\end{equation}
where $\rho_i$ is the volume density and $l_i$ is the length of each material.  
The constant of proportionality depends on detector acceptance, but since the target materials 
do not change with time, this constant is the same for the numerator and 
denominator, and it cancels out of the ratio. 
The dilution factor for the $^{14}$NH$_3$ target is then 
(we replace the symbol $^{14}$NH$_3$ by a symbol $A$) :
\begin{equation}
\small
\frac{ \frac{3}{17} {\rho_A l_A  \sigma_A}} 
{ {\rho_{He}(L-l_A)\sigma_{He}}
 +{ (\frac{14}{17}\sigma_N+\frac{3}{17}\sigma_p)\rho_A l_A}
 +{\rho_{Al}l_{Al}\sigma_{Al}}
 +{\rho_K l_K \sigma_K}
},
\label{eq:dilFacExample} 
\end{equation}

\noindent where $L$ is the length of the He bath
and $l_A$ is the effective ammonia length.
The radiated cross sections are a function of the 
length of the material in units of radiation length 
and are obtained by modeling the available world data, 
and the areal density for each material was measured 
in the lab or obtained from the literature.
The average areal densities of each of the 
materials are listed in Table \ref{table1}.

\begin{table}[h]
\begin{tabular}{lcccccc}
       & H         &    D      &   He      &   C       &   N   & Al \\
Target &           &           &           &           &       & \\
\hline
NH$_3$ &   0.133 &   0      &   0.164 &   0.009 &   0.621 &   0.045 \\
ND$_3$ &   0.015 &   0.239  &   0.161 &   0.009 &   0.628 &   0.045 \\
Empty  &   0     &   0      &   0.290 &   0.009 &   0     &   0.045 \\
Carbon &   0     &   0      &   0.232 &   0.886 &   0     &   0.045 \\
\hline
\end{tabular}
\caption{Average areal densities in units of 
g/cm$^2$ for each of the materials in the four
targets of the experiment. Kapton is not 
mentioned here, but is approximated as pure carbon.
The results correspond to $L=2.0$ cm and $l_A=0.87$ cm 
($l_A=0.89$ cm) for the NH$_3$ (ND$_3$) target.
}
\label{table1}
\end{table}

As shown by Eq.~\ref{eq:dilFacExample}, to accurately determine the dilution factor 
we need to know the total length of the He bath ($L$) in the target and the packing fraction 
or `length'  ($l_A$) of the NH$_3$ beads in the target cup.

\subsubsection{Length of the He bath ($L$)}\label{sec:targetLength}
The distance $L$ between the aluminum windows that contain the liquid He was measured at room temperature to be 
2.3~$\pm$~0.3~cm.  The large uncertainty arises because the windows are very thin and 
hence can bow inwards or outwards, depending on pressure differences.We therefore relied primarily 
on data taken with the empty target and 
no helium bath, where two peaks in the vertex-$Z$ distribution
could be clearly seen, corresponding to the windows.
The peak separation from these measurements
is 2.0~$\pm$~0.1~cm.
	
A check was made by extracting $L$ from measurements of the empty and carbon targets,
with and without helium present, and assuming a helium density of 0.145 
g/cm$^2$.The average value of $L$ from a set of 
ten such measurements was $2.1\pm0.1$ cm.

\subsubsection{Ammonia length $l_A$ (the packing fraction)}
The other quantity extracted is the packing fraction of the ammonia beads in the Kapton target cup. 
This gives us the apparent length of the ammonia target  as if it were packed into a solid 
piece as opposed to crushed beads.

The method used to determine $l_A$ was to fit for the value that best described the measured
ratios of electron rates from the carbon and ammonia targets.  The predictions used fits to world
data taking into account radiative effects using the Mo-Tsai formalism ~\cite{motsai}.  The dependence
of external radiative corrections on $l_A$ were taken into account in the fits.
For each carbon target run, the run on ammonia closest in time was chosen to obtain a value of $l_A$
in order to minimize the effects of slow drifts in average detection 
efficiency.
A summary of the results is presented in Table \ref{table2}.
	
\begin{table}[h]
\centering
\begin{tabular}{lll}
Part & Target Cup & Packing fraction $l_A$ \\ 
\hline
Part A    & Top    & $0.85\pm 0.03$  cm \\ 
Part A    & Bottom & $0.85\pm 0.03$  cm \\ 
 
\hline
Part B    & Top    & $0.86\pm 0.03$   cm \\
Part B    & Bottom & $0.91\pm 0.03$  cm \\
\hline
Part C    & Top    & $0.92\pm 0.03$   cm \\ 
Part C    & ND3    & $0.89\pm 0.025$   cm \\
\end{tabular}
\caption{Effective ammonia
lengths,$l_A$, for 3 run periods and 2 target cells. 
}
\label{table2}
\end{table}

\subsubsection{NH$_3$ Contamination of the ND$_3$ Target}
\begin{figure}[hbt]
\centerline{\includegraphics[width=3.2in]{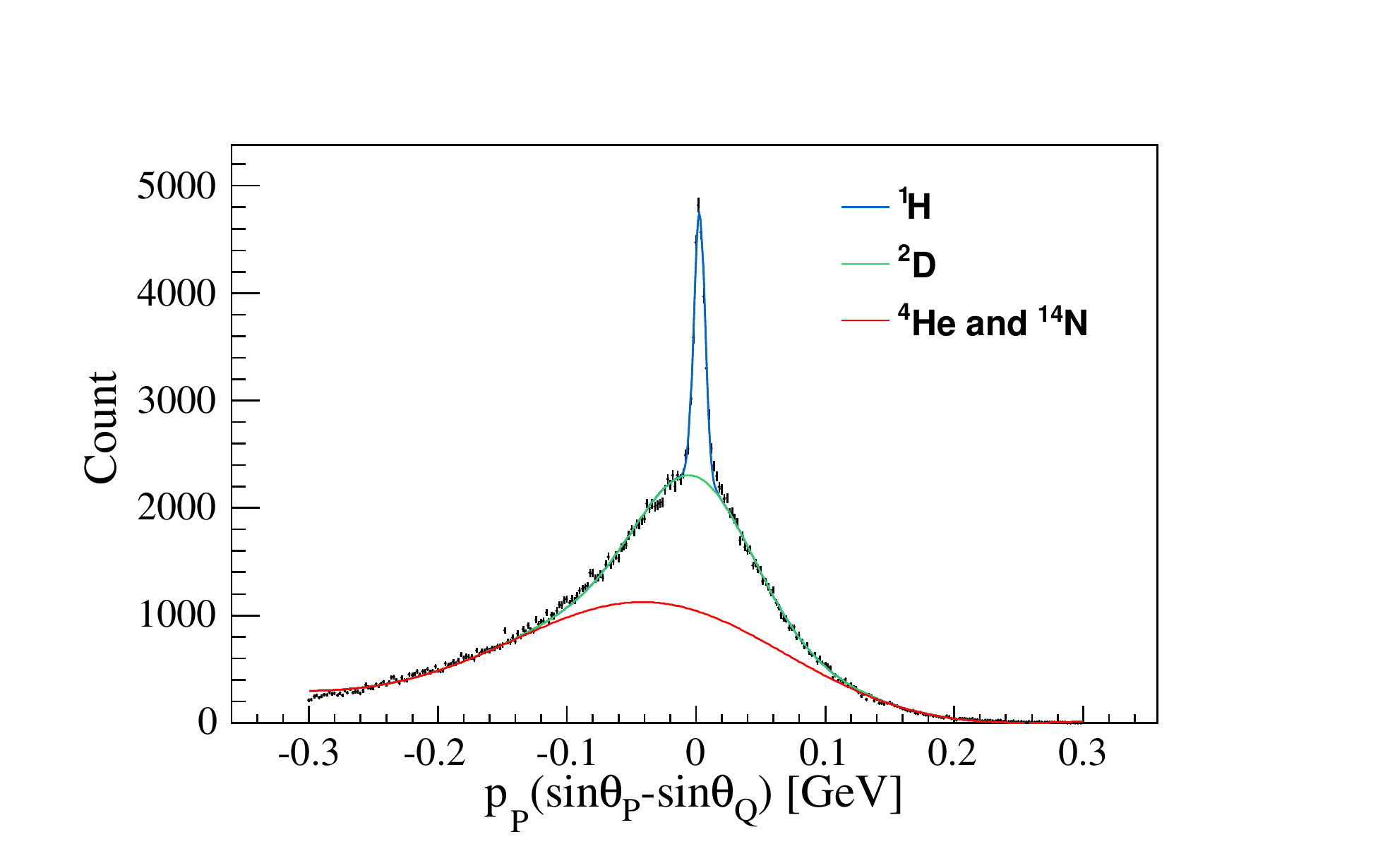}}
\caption{(color online) Distribution of $p_T^q$ for exclusive $ep$ elastic
scattering from the ND$_3$ target.
}
\label{fig:contam3}
\end{figure}

The manufacturer's specification for the deuterium gas used
to make the ammonia beads had $<1\%$ hydrogen contamination.
A check was made using $ep$ elastic data, and the actual
contamination was, surprisingly, found to be an order
of magnitude larger. After standard $ep$ elastic
exclusivity cuts, the events from hydrogen, deuterium,
and heavier nuclei could clearly be distinguished by
plotting the event rates as a function of missing
transverse momentum, $p_T^q$,  relative to the momentum transfer
vector $\vec q$. In the case of a free proton,
the  $p_T^q$ distribution is a delta-function
widened by our experimental resolution. In the case
of the deuteron, it is considerably wider due to the
average 50 MeV Fermi motion of the proton in the deuteron.
In the case of heavier nuclei such as carbon and nitrogen,
the peak is another factor of four wider, because the typical
Fermi momentum is on the order of 200 MeV. These features are
illustrated in Figure~\ref{fig:contam3}.
	
	The distribution of $p_T^q$
for the carbon target is very wide, and that
for the NH$_3$ target has a very narrow peak, 
sitting on top of a wide distribution with the same 
shape as for the carbon target. The ND$_3$ target spectrum
has the expected medium-width peak from free deuterons
in the target, sitting on top of a nuclear 
background from nitrogen. Unfortunately, a very narrow
peak of the same width as seen in the NH$_3$ target
is also clearly visible. Using data from the carbon
and NH$_3$ targets as a guide, we performed fits to 
the three components visible in the ND$_3$ spectra
to obtain the relative fraction of free protons and
deuterons. The result of the study was that 
$10.5\pm0.4\%$ of the ND$_3$ effective
target length was NH$_3$, for the in-bending runs, and
$12.0\pm0.7\%$ for the outbending runs. No time-dependence
to the contamination was observed within each of these
run periods.

\subsubsection{Dilution factor results}
\begin{figure}[ht]
\centerline{\includegraphics[width=3.2in,angle=90]{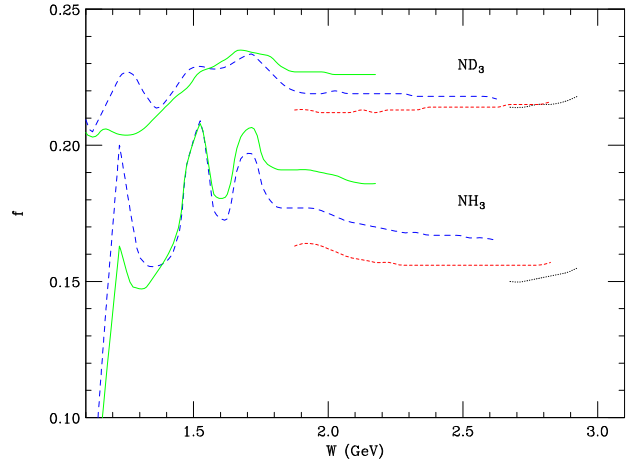}}
\caption{(color online) Dilution factor as a function of $W$ for 
$Q^2=1$ GeV$^2$ (black, dotted curve), 
$Q^2=1.6$ GeV$^2$ (red, short-dash curve), 
$Q^2=3$ GeV$^2$ (blue, long-dash curve), and
$Q^2=5$ GeV$^2$ (green, solid curve).
The lower (upper) sets of curves are for the NH$_3$
(ND$_3$) target.
}	
\label{fig:dfresult}
\end{figure}

Having determined the areal density of 
each component of each target, we used the
 method outlined above to calculate the 
corresponding dilution factor as a function of $(x, Q^2)$
and $(W,Q^2)$. The results for NH$_3$, using a typical value
$l_A=0.87$ cm, are shown in Fig.~\ref{fig:dfresult} as a function
of $W$ for four values of $Q^2$. 

Clear resonance structure can 
be seen corresponding to the three well-established resonance
regions near $W=1.23$, 1.5, and 1.7 GeV, arising primarily
from the $W$-dependence of the ratio of proton to neutron cross
sections. The dilution factor for the deuteron shows much weaker structure,
because the deuteron and the $A>2$ materials both contain almost
equal numbers of proton and neutrons.

\subsection{$c_s$ Correction}
The $c_s$  correction factor primarily takes into account the background
contribution from electrons produced in $\pi^0$ decays. 
The $c_s$ correction also accounts for a smaller background, from
$\pi^-$ mesons misidentified as electrons. 

\begin{figure}[hbt]
\centerline{\includegraphics[width=3.2in] {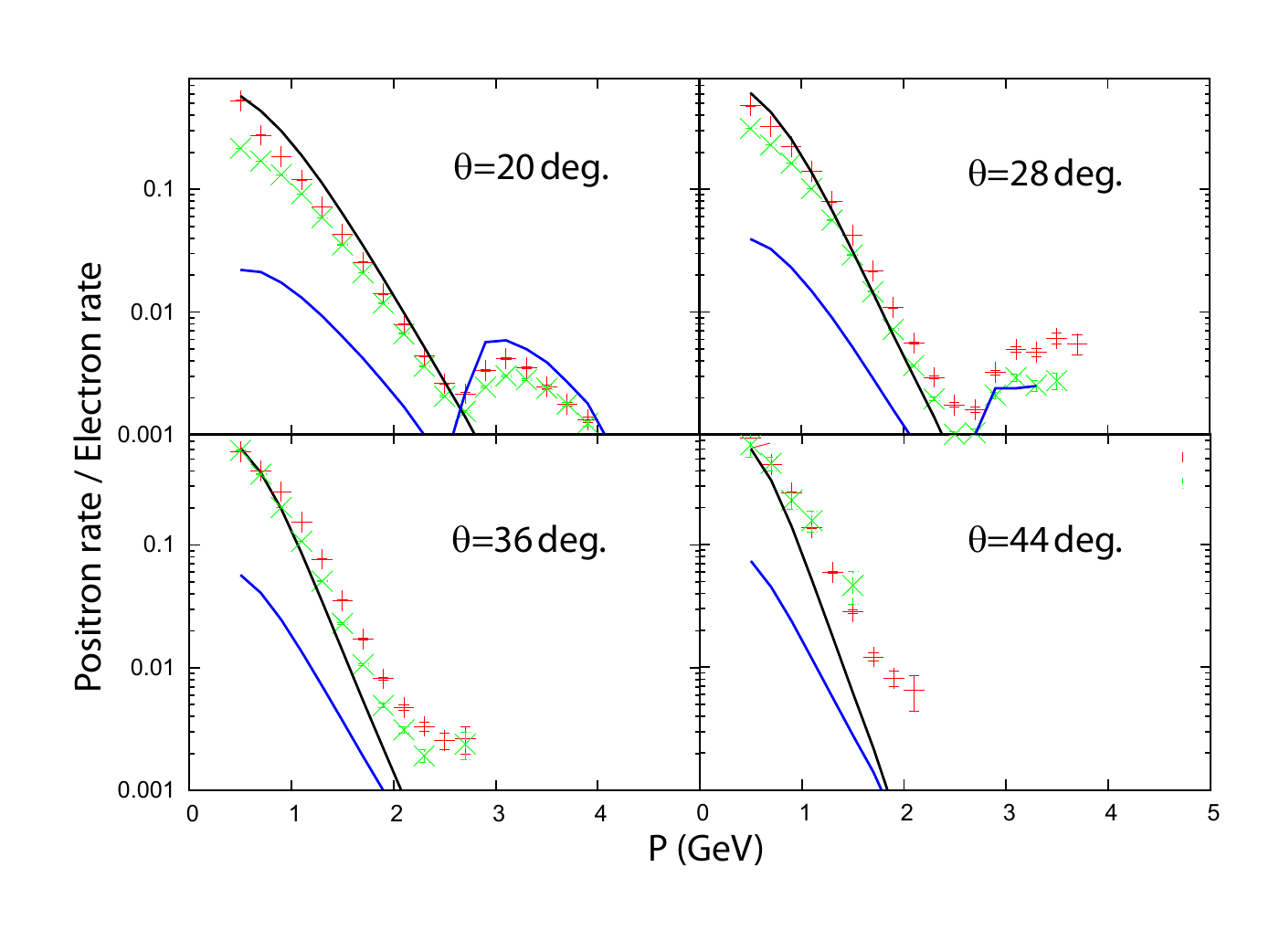}}
\caption{(color online) Ratio of positron to electron 
inclusive rates ($c_s-1$) as a function of momentum for four
bins in polar scattering angle $\theta$.
The green points are for both particles out-bending in the
torus and the red points are for both particles in-bending.
The curves are from a simulation (see text for details).
}
\label{fig:posRate2x2}
\end{figure}

Dalitz decay of the $\pi^0$ 
($\pi^0 \rightarrow \gamma e^+e^-$)
or Bethe-Heitler conversions of one of the photons
from ``normal'' $\pi^0$ decay 
($\pi^0 \rightarrow \gamma \gamma$)
both produce electron-positron pairs. Since there are equal numbers of electrons and
positrons produced from $\pi^0$ decays, the rate
and asymmetry of positrons can be used to correct
the electron sample. We therefore define $c_s= 1 + (N_{e^+}/N_{e^-})$.
To compare the electron/positron rates 
within the same acceptance, with both particles in-bending or both particles
out-bending, we use our data taken with opposite torus fields.
The results, binned in momentum and 
polar angle ($p$, $\theta$), 
are shown in Figure~\ref{fig:posRate2x2}, for the
case where both electron and positron are in-bending
(red), and where both are out-bending (green). The average of
the two was used to evaluate $c_s$.
It can be seen that the ratios are negligibly small for $P>2$ GeV,
independent of $\theta$, but rise to very large values 
below $P=1.2$ GeV. We therefore used the cut $P>1.2$ GeV,
corresponding to $y =\frac{\nu}{E}<0.8$, in this analysis.

The ratios of $e^+/e^-$ are compared with simulations based on
the known rates of pion yields with 6 GeV electron beams~\cite{Wiser}.
The black curves are from $\pi^0$ decays, while the blue
curves are from mis-identified $\pi^-$.
The simulation accounts quite well for the magnitude and
kinematic dependence of the observed ratios. The 
enhancement seen above 2.65 GeV is a consequence of this
value being the threshold for charged pions to produce light
in the gas Cherenkov counter.

The functional form of the $c_s$ correction (basically a dilution
factor), arises from the assumption that the double-spin
asymmetry for inclusive pion production is zero. Our 
measurements of positron asymmetries were found to be 
consistent with this assumption, when
the positron asymmetry is at least one order of magnitude
smaller than the electron asymmetry.

\subsection{Polarized Nitrogen Contribution}\label{sec:c1}
The $c_1$ term in Eq.~\ref{Eq:Aparrp}
 accounts for the polarized 
nitrogen contribution to the measured 
double spin asymmetry. From the definition
of the raw asymmetry in terms of the physics
asymmetries of each of the polarizable nuclei
in a target, it is straightforward to show that
\be
c_1 = \frac{\eta_N}{\eta_{p,d}} 
      \frac{A_N\sigma_N}{A_{p,d}\sigma_{p,d}}
      \frac{P_N}{P_{p,d}}, 
\label{Eq:c1}
\ee
where $\eta$ is the number of nuclei of a given species,
$\sigma$ is the cross section per nucleus, 
$A$ is the double-spin asymmetry (hence $A\sigma$
is the cross section difference), and $P$ is
the polarization of a given material. For each
of these four variables, the subscript $N$ is 
for nitrogen-14, and 
$(p,d)$ refers to either the proton (for the NH$_3$
target) or deuteron (for the ND$_3$ target). 
The first term $(\eta_N / \eta_{p,d})$ 
in Eq.~\ref{Eq:c1} is 1/3, by definition, for
ammonia.

In the nuclear shell model, the spin-1 nitrogen-14 
nucleus can be considered
as a spinless carbon nucleus surrounded by
an extra proton and neutron, each in a 
$1p_{\frac{1}{2}}$ orbital state~\cite{Rondon}. 
After doing
the spin projections, it turns out that the proton
and neutron are each twice as likely to have their
spin anti-aligned with the nitrogen spin, as having
it aligned. It then follows that the second
factor in Eq.~\ref{Eq:c1} can be evaluated using
\be
A_N\sigma_N = -\frac{1}{3} (\sigma_p A_p + \sigma_n A_n) =
-\frac{1}{3}\sigma_d A_d,
\ee
where the subscript $n$ refers to the neutron,
and we have neglected the small d-state correction
and used the relation $\sigma_d = \sigma_p + \sigma_n$.
Inserting this into the second term of  Eq.~\ref{Eq:c1},
we obtain a constant value of  
$-0.33\pm 0.08$ for the ND$_3$ target
and $(-0.33\pm 0.08) A_d/A_p$ for the NH$_3$ target.
The uncertainty of 0.08 comes from an 
evaluation~\cite{SMCNMR} of a range of more
sophisticated treatments of the nitrogen wave
function than the simple shell model.

The third factor, the ratio of nitrogen to 
proton (deuteron) polarizations,
can be evaluated using Equal Spin Temperature (EST)
theory~\cite{EST,SMCNMR}. This gives $p_N/P_p=0.098$
for the average value $P_p=0.77$ of this experiment.
Experimental measurements from the SMC~\cite{SMCNMR}
are consistent with this result, although about
10\% to 15\% higher. We therefore used
$P_N/P_p=0.10\pm0.01$.
The EST theory predicts $P_N/P_d=0.48$,
essentially independent of $P_d$. An experimental
study at SLAC (E143 experiment, unpublished) yielded a 
much lower value of $P_N/P_d=0.33$. We therefore used
an average: $P_N/P_d=0.40\pm0.08$. 

Combining all these results together yields:
\be
c_1^p = (-0.011\pm 0.003) (\sigma_d/\sigma_p) (A_d/A_p) 
\ee
\be
c_1^d = -0.044 \pm 0.014.
\ee
To evaluate $\sigma_d/\sigma_p$, the ratio of deuteron
to proton cross section per nucleus, we used recent
fits to world data~\cite{christy,bosted}. To
evaluate the ratio of double-spin asymmetries
$A_d/A_p$ we used an empirical  fit to previous 
world data~\cite{Eg1bLONGPAPER}.
 
The values of $c_1^p$ vary between -0.007 at low $x$ and -0.013 at high $x$,
with little dependence on $Q^2$.

\subsection{Correction for NH$_3$ in ND$_3$}
The $c_2$ term in Eq.~\ref{Eq:Aparrd}
 accounts for the polarized NH$_3$ 
contribution to the measured 
double spin asymmetry in the nominal ND$_3$ target.
As discussed above, the ND$_3$ target contained
an approximately 10.5\% (by weight) 
admixture of NH$_3$ (or
equivalent), and the protons in this material
were polarized. In the following treatment, we consider 
NH$_3$ as a background, and we define $c_2$ as

\be
c_2 = \frac{\eta_p}{\eta_d} \frac{P_p}{P_d} 
\label{Eq:c2},
\ee
where the variables have the same meaning as in
Eq.~\ref{Eq:c1}. The derivation of Eq.~\ref{Eq:c2}
is valid only if the dilution factor $f$ in 
Eq.~\ref{Eq:Aparrd} is defined using the number of
polarizable nucleons in deuterium only (not including
the free protons), in the numerator of the ratio.

From the discussion above, the ratio of proton to
deuteron nuclei is $0.105\pm0.004$ ($0.120\pm 0.006$)
for the in-bending (out-bending)
portion of Part C. The ratio of proton to deuteron
polarizations is $2.31\pm0.2$ ($2.15\pm0.3$) for
the in-bending (out-bending) portion of Part C. 
Numerically, we then obtain for the average values:
\be
c_2 = (0.24\pm0.02) \hskip 0.1in  \hskip 0.5in {\rm in-bending}
\ee
\be
c_2 = (0.26\pm0.04)  \hskip 0.1in \hskip 0.5in {\rm out-bending}.
\ee


\subsection{Radiative Corrections}
In our analysis we approximate the scattering process 
as a one-photon exchange process, also called Born scattering. The higher order 
processes contributing to the total measured cross sections and asymmetries are 
taken into account by calculating radiative corrections. 
The radiative corrections can be broken into two kinds: internal and external. 
The internal processes occur within the field of the scattering nucleus 
and consist of the vertex corrections [which effectively account for
the running of the fine coupling ``constant'' $\alpha(Q^2)$], as well
as the emission of hard photons from the incident or scattered electron.
External radiation occurs when a Bremsstrahlung photon
is emitted from the incident electron prior to scattering (from a different
nucleus from which the hard scattering takes place), or a hard photon
is radiated by the scattered electron.
The probability of emitting a hard photon is approximately given by $t dk/k$, where $t$ is
the material thickness in radiation lengths, and $k$ is the photon
energy. An important consideration is
that an electron is de-polarized by the emission of Bremsstrahlung
photons. As a rough guide at JLab energies, internal radiation is equivalent to
external radiation when $t$ is on the order a few percent. 
The main difference between the two is that the
electron angles are essentially unchanged in external radiation
(characteristic angle $m_e/E$), whereas significant changes in the
electron scattering at the vertex can occasionally occur in
the internal radiation process. For our ammonia targets, the
values of $t$ relevant for external radiation are about 1.2\% 
for the incident electron and 4\% for the scattered electron.

The radiative corrections require the evaluation of 
both polarized and unpolarized components for Born, 
internally radiated, and fully radiated cross sections and asymmetries.
Polarization-dependent internal radiative 
cross sections were calculated using the 
formalism developed by Kuchto and Shumeiko \cite{kuchto,E143}. External
radiation was taken into account by convoluting internal radiative
corrections with the spectrum of incident electron energies,
rather than a single monochromatic value, according to the
formalism of Mo and Tsai~\cite{motsai}. The external radiation from
the scattered electron was similarly taken into account. 

The calculations were done with the computer code
RCSLACPOL, developed for the E143 experiment at SLAC
in the early 1990's~\cite{E143}. The code requires input
models for inelastic electron scattering as well as $ep$ elastic
(proton target) or $ep$ and $en$ quasi-elastic scattering. 
For spin-averaged inelastic cross sections, we used the recent
fits to world data of Christy and Bosted~\cite{christy} for the
proton and Bosted and Christy~\cite{bosted} for the deuteron.
Spin-dependent inelastic cross sections were obtained using a
recent fit to JLab data~\cite{nguler}. 

In our correction scheme, the radiative corrections are broken into 
an additive correction $A_{RC}$, and 
a radiative dilution factor, $f_{RC}$. 
The factor $f_{RC}$ is nothing more
than ($1-f_e$), where $f_e$ is 
the fraction of events that have radiated down into
a given $(x,Q^2)$ bin from the $ep$ elastic scattering
process (or quasi-elastic process for the deuteron target).
The factor $A_{RC}$ accounts for all other radiative
processes.
The radiatively corrected asymmetry is then given by 
\begin{equation}
A_{corr}=A_{uncorr}/f_{RC}-A_{RC}.
\end{equation}
The statistical uncertainty on the corrected asymmetry is given by
\begin{equation}
\delta A_{corr} = \delta A_{uncorr}/f_{RC}.
\end{equation}

The corrections were found to be 
very small for $y=\frac{\nu}{E} < 0.8$, the cut used in this analysis, corresponding to typical changes in
$g_1/F_1$ of less than 1\%. The largest effects are in 
the resonance region, where the asymmetry is changing
rapidly with $W$. 
The error bars with
radiative corrections applied are larger than
without corrections, with the biggest increase at
the largest $W$ of a given $Q^2$ bin (corresponding
to large values of $y$). The effect of radiative
corrections for the deuteron is even smaller than
for the proton.

\section{RESULTS AND COMPARISON TO THEORY}\label{s5}

\subsection{Results for $g_1 / F_1$}
In the one-photon-exchange (Born) approximation, the cross section for inclusive
electron scattering with beam and target spin parallel ($\uparrow\Uparrow$) 
or anti-parallel ($\uparrow\Downarrow$)  
can be expressed in terms of the four structure functions 
$F_1, F_2, g_1$ and $g_2$, all of which can 
depend on $\nu$ and $Q^2$\cite{Eg1bLONGPAPER}:
\begin{eqnarray}
\label{sigfull}
\frac{d\sigma^{\uparrow\Downarrow/\uparrow\Uparrow}}{d\Omega dE'} & = &
 \sigma_M \left[ \frac{F_2}{\nu} + 2\tan^2\frac{\theta}{2} \, \frac{F_1}{M} \pm 2\tan^2\frac{\theta}{2} \right. \nonumber \\
 & &  \left.  \times \left( \frac{E_0 + E' \cos\theta}{M \nu} g_1 - \frac{Q^2}{M \nu^2} g_2 \right) \right] ,
\end{eqnarray}
where the Mott cross section
\begin{equation}
\label{sigmott}
\sigma_M = \frac{4E'^2 \alpha^2 \cos^2\frac{\theta}{2}}{Q^4} .
\end{equation}
We can now define the double spin asymmetry $A_\|$ as
\begin{equation}
\label{Aparintro:eq}
A_\|(\nu,Q^2,y) = \frac{d\sigma^{\uparrow\Downarrow} - d\sigma^{\uparrow\Uparrow}}{d\sigma^{\uparrow\Downarrow} + 
d\sigma^{\uparrow\Uparrow}} .
\end{equation}
The measured asymmetry
$A_\|$ contains contributions from both the $g_1$
and $g_2$ structure functions. After some algebra,
the equations can be rewritten as:
\begin{equation}
g_1/F_1 = (A_\| /D^\prime) C_{g_2},
\label{Eq:g1f1}
\end{equation}
\noindent where the depolarization factor in this
case is given by~\cite{E143}
\begin{equation}
D^{\prime} = \frac{(1 -\epsilon)(2-y)}{y(1 +\epsilon  R)},
\label{Eq:Dprime}
\end{equation}
with $\epsilon = (1+2(1+\tau)(\tan(\frac{\theta}{2}))^2)^{-1}$ and $\tau = \frac{\nu^2}{Q^2}$.  
The ratio $R$ of longitudinal to transverse virtual photon absorption cross section is:
\begin{equation}
\label{Rsig}
R = \frac{\sigma_L({\gamma^*})}{\sigma_T({\gamma^*})} = 
\frac{F_2}{2 x F_1} (1+\gamma^2) - 1,
\end{equation}
where $\gamma = \frac{\sqrt{Q^2}}{\nu}$.
The correction factor due to $g_2$, $C_{g_2}$ is described in the section below.

\subsubsection{Depolarization Factor and R}
	
This depolarization factor is a function of $R$ (Eq.~\ref{Rsig}),
which we obtained from a fit to world data~\cite{christy}.  
The fit describes the range 
$0 \le Q^2 \le 8 $ GeV$^2$ and $1.1 \le W \le 3.1 $  GeV which 
covers the area of interest for this study. Typical 
values of $R$ vary between 0.1 and 0.3 for both the 
proton and the deuteron, with a typical uncertainty of 0.03.

\subsubsection{$g_2$ Correction}
\begin{figure}[hbt]
\centerline{\includegraphics[width=5in,angle=90]{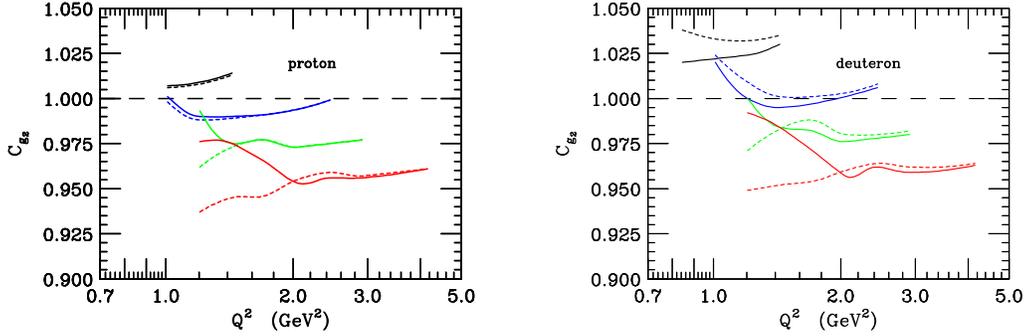}}
\vskip -2.5in
\caption{(Color online) Correction factor $C_{g_2}$ as a function of
$Q^2$. The curves are, from top to bottom, $x=0.225$ (black), 
$x=0.325$ (blue), 
$x=0.425$ (green), and $x=0.525$ (red). Left panel
is for the proton target, right panel for the deuteron.
Dashed curves use the lowest-twist Wandzura and Wilczeck
formula, while the solid curves use a fit to world
data.}
\label{fig:g2corr}
\end{figure}

 The $g_2$ correction factor is given by:
\begin{equation}
\small
C_{g_2} = \frac{1 + E^\prime/E_0}
{1+E^\prime\cos(\theta)/E_0} \hskip .1in 
\frac{1}{1 - 2(g_2/g_1) M x/[E_0 + E^\prime\cos(\theta)]}.
\end{equation}
To gauge the rough order of magnitude of the correction,
we note that for $\cos(\theta)=1$, $x=0.5$, $E_0=6$ GeV, then
\begin{equation}
C_{g_2} \approx 1 + (g_2/g_1)/10.
\end{equation}
Since $g_2$ is smaller in magnitude than $g_1$,
the overall correction differs by only a few percent
from unity, as illustrated in Fig.~\ref{fig:g2corr}.

The values of $g_2/g_1$ used in the analysis were
taken from a fit to world data~\cite{nguler}. In
order to estimate the systematic uncertainty on the correction,
we also used the assumption that there are no
deviations from the twist-two model of Wandzura
and  Wilczeck ($g_2^{WW}$) \cite{WW}, in which
\begin{equation} 
g_2(x,Q^2) = -g_1(x,Q^2)
+\int_x^1 g_1(\xi,Q^2)d\xi/\xi. 
\end{equation}
In this relation, the magnitude of $g_1$ cancels, and
the result depends only on the $x$-dependence at
higher $x$. The corrections $C_{g_2}$ 
using only the Wandzura and Wilczeck contribution
are shown as the dashed curves in Fig.~\ref{fig:g2corr}.  
The differences from the world
fit of Ref.~\cite{nguler} are quite small, except for $x=0.525$ at 
low $Q^2$.

\subsubsection{Combining Data Sets}
    Physics quantities that should depend only on
    $(x,Q^2)$ [or equivalently $(W,Q^2)$] were first
    calculated for each beam energy and torus polarity.
    These physics quantities were then combined,
    weighted by their statistical uncertainties. The proton
    results used only Parts A and B, due to the very
    small amount of proton data in Part C. The deuteron
    data are from Part C only. Data from each of the
    individual run periods were compared with the
    corresponding averages, and found to be consistent
    within overall normalization uncertainties (dominated
    by the uncertainty in $P_bP_t$).

\subsection{Systematic Uncertainties}
In this section we first summarize the systematic
uncertainties on $A_\|$ arising from each of the terms
in Eqs.~\ref{Eq:Aparrp} and ~\ref{Eq:Aparrd}. We then discuss the systematic
uncertainties from three other additional sources, which
were assumed to be negligible in these equations.

\subsubsection{Beam charge ratio $r_c$}
The uncertainty in the ratio of incident beam charge for
positive and negative helicities (relative to the target
polarization direction) was much less than 0.0001 on a few-minute 
time scale. This
negligibly small value was achieved by three methods: 1) frequent
reversal of the half-wave plate in the CEBAF injector; 2) keeping the charge
asymmetry less than 0.1\% using an on-line feedback system;
and 3) generating beam helicity buckets in pairs. 

\subsubsection{Dilution factor $f$}
The dilution factor is one of the two most important 
sources of systematic uncertainties in the determination of $A_\|$.
There are several factors which contribute here:
\begin{itemize}
\item
The ammonia length $l_A$ has an estimated relative
uncertainty of 3\% for most of the experiment, based
on the spread in the individual determinations for
pairs of carbon and ammonia runs, combined with
an overall uncertainty of 1\% in the carbon target
areal density.
\item
The length of the He bath, $L$ (which
determines how much helium there is in the target), has
an estimated uncertainty of 0.2 cm, based on inconsistencies
between the determination of 2.0 cm from the empty target
runs, and 2.3 cm for a direct measurement in the lab
when the target was at room temperature. 
\item
The areal density of the aluminum He containment
windows has an estimated uncertainty of 0.005 g/cm$^2$
(a relative uncertainty of 10\%).
\item
The areal density of the target Kapton windows has
an estimated uncertainty of 0.005 g/cm$^2$ (a 
relative uncertainty of 5\%)
\item 
Approximately 3\% uncertainties in the
density of ammonia and helium at 1 K.  
\end{itemize}
Folding together these uncertainties
for our particular target (i.e. about 70\% ammonia
by areal density, 30\% other nuclei), the result
is a 1.5\% relative uncertainty in $f$, with no significant
$(x,Q^2)$ dependence for $W>1.4$ GeV. 

Although all of the
above contributions vanish in the limit of a pure
ammonia target, there is still the overall scale
uncertainty in the ratios 
of spin-averaged inclusive cross sections $\sigma_p/\sigma_{14N}$
and $\sigma_d/\sigma_{14N}$. Based on the fluctuations
between various experiments fit in 
Refs.~\cite{bosted,christy,Adep}, we estimate the
uncertainties be 1.5\% in both cases. Combining this
with the target-parameter-dependent uncertainty of 1.5\%,
the total relative uncertainty in $f$ is 2.3\% for both proton and deuteron.

\subsubsection{Product of beam and target polarization 
$P_BP_T$}
For the proton, the relative uncertainty on $P_bP_t$ has a statistical
component of 1\% (averaged over Parts A and B). We
estimate a further systematic uncertainty of 1\% coming from
the choice of cuts and the model for $G_E/G_M$. The net
systematic uncertainty on $P_bP_t$ for the proton is therefore
1.4\%.

For the deuteron, we estimate an overall normalization
uncertainty of 3\% to 5\% in the target polarization values
extracted from the NMR signals by the double-peak method.
The systematic uncertainty on the beam polarization from 
the M\o ller methods is estimated to be 4\%, based on
comparison with more accurate measurements made in
Halls A and C over a period of many years.
The net result is an overall relative systematic uncertainty of 
7\% in $P_bP_t$ for the deuteron.

\subsubsection{Uncertainty associated with pair-symmetric and mis-identified pion correction}
The correction term, $c_s-1$,is
negligibly small for $P>2$ GeV ($y<0.65$), but rises to
values as large as $c_s-1=0.10$ at $P=1.2$ ($y=0.8$), the
lowest value of $P$ used in the analysis. Based on
the disagreement between our two measurements (both
particles in-bending, or both out-bending), we assigned
a systematic uncertainty of 30\% on $c_s-1$, corresponding to
a systematic uncertainty of up to 3\% on $A_\|$ at the
highest values of $y$.

We assumed that the pair-symmetric asymmetry was zero, however,
at low $P$, there are some indications that the
scaled contribution could be as large as 1\% for
$1.2<P<1.5$ GeV for the proton, and as much as 
3\% for the deuteron. We therefore assigned a relative
systematic uncertainty of 1\% (3\%) to $A_\|$ for the proton
(deuteron) for $P<1.5$ GeV. 

\subsubsection{Uncertainty associated with radiative corrections}
The application
of radiative corrections make typically less than 1\%
changes to $g_1/F_1$. To study the systematic uncertainty,
radiative corrections were calculated with several
alternate cross section and asymmetry models. No
significant changes were observed, at the 0.5\% level.

\subsubsection{Uncertainty on $c_1$}
The uncertainty on the value of $c_1$ was estimated to be 0.003 (0.014) for the proton (deuteron), 
independent of $(x,Q^2)$, based on the uncertainties in
modeling nitrogen-14, as discussed in Sec.~\ref{sec:c1}.

\subsubsection{Uncertainty on $c_2$}
The uncertainty on the value of $c_2$ was estimated
to be 0.025 (0.038) for the in-bending (out-bending) deuteron asymmetries, 
independent of $(x,Q^2)$. The corresponding relative uncertainty in the deuteron 
asymmetries and structure functions is given by $0.025  A_\|^p/A_\|^d$ for the 
in-bending runs, and $0.038 A_\|^p/A_\|^d$ for the out-bending runs, where the ratio
 $A_\|^p/A_\|^d$ varies between 2 and 3 over the $(x,Q^2)$ range of the experiment.

\subsubsection{Detection Efficiency}\label{sec:luminosity}
In Eqs.~\ref{Eq:Aparrp} and \ref{Eq:Aparrd}, we assumed that the 
detector efficiency was the same for target and beam
polarization aligned or anti-aligned. 
A correction should be made if the overall
particle  rate is higher for one
state than the other, resulting in a rate-dependence
to the detection efficiency. We estimated that
if the total particle rate were entirely good electrons, the
measured asymmetry would be 7\% lower than if the
detector efficiency were not rate-dependent. Fortunately, the 
total particle rate was dominated ($>95\%$) by photo-produced
pions, rather than scattered electrons.

From measurements at SLAC~\cite{E155pion}, we
know that the raw asymmetry in pion production is
an order-of-magnitude smaller than for electron scattering.
Therefore the correction would be on the order of 0.7\%.
Taking into account that approximately the same correction
applies to $ep$ elastic scattering (from which we determine
$P_BP_T$), the net effect on the final electron
asymmetry is reduced even further. We therefore made
no correction for rate-dependent detection efficiency,
and  assign an overall systematic uncertainty of 0.7\%.

\subsubsection{Parity-violating background}
The raw asymmetry arising due to
$eN$ parity-violating inelastic
scattering from any of the nucleons in the target
is given to a good approximation by
$A_\|^{PV} = P_b Q^2 [0.8 \times 10^{-4}]$, independent 
of $x$ (from the Particle Data Book~\cite{pdg}). Since $A_\|^{PV}$ does not depend on target
polarization $P_T$ (unlike the double spin asymmetry
$A_\|$), the contribution to $A_\|$ cancels, by definition,  for equal 
running times with the target
polarization aligned (denoted by $t^+$)
or anti-aligned ($t^-$) with the beam direction.
For unequal running times, the contribution from $A_\|^{PV}$ is reduced
by a factor $r_T=(t^+-t^-)/(t^+ + t^-)$.  
Averaged over the entire experiment,
$r_T=0.04$ for NH$_3$ and $r_T=0.11$ for ND$_3$.
Since the measured raw double-spin raw asymmetry is
approximately given by $Q^2 \times 10^{-2}$ for
our average virtual photon energy $\nu=3$ GeV, 
the relative parity-violating
contribution was less than 0.1\% in all $(x,Q^2)$ bins,
and was neglected.
	
\subsubsection{Summary of systematic uncertainties on $A_\|$}
With all of the above uncertainties added in quadrature, the
overall relative systematic uncertainty on $A_\|$ has 
an $(x,Q^2)$-independent value of 2.8\% for the
proton and 8.1\% for the deuteron. The proton
uncertainty is dominated by the uncertainty in $f$, while
the deuteron is dominated by the uncertainty in $P_bP_t$.
The only
systematic uncertainty that depends strongly on kinematic
values for the proton is the pair-symmetric correction, which increases
the overall uncertainty to 4.1\% (8.6\%) for the proton
(deuteron) at $y=0.8$. For the deuteron, the $c_2$ contribution 
varies significantly with kinematics also.
A summary of systematic uncertainty factors contributing to  $\Delta A_\|$
is presented in Table \ref{systerrtable}.
	
\begin{table}[hbt]
\centering
\begin{tabular}{l|c|c}
\multicolumn{1}{c}{Factor}
& \multicolumn{1}{c}{Proton}
& \multicolumn{1}{c}{Deuteron} \\
\hline
$f$          & 2.3\%      &  2.3\%    \\ 
$P_bP_t$     & 1.4\%      &  7.0\%    \\ 
$c_s$        & 0-3\%      &  0-3\%    \\
$c_1$        & 0.3\%      &  1.4\%  \\
$c_2$        & n/a        &  5-10\%    \\
$R.C.$       & 0.5\%      &  0.5\%     \\
$r_c$        & 0.1\%      &  0.1\%     \\   
\hline
total        & $2.8-4.1\%$ & $8.1-12\%$ \\
\end{tabular}
\caption{Summary of relative systematic uncertainties 
in $A_\|$ for proton and deuteron.
}
\label{systerrtable}
\end{table}

\subsection{Systematic Uncertainty on $g_1/F_1$}
Eq.~\ref{Eq:g1f1} shows that the systematic
uncertainty on $g_1/F_1$ has two additional sources 
compared to $A_\|$: the uncertainty on $D^\prime$
(dominated by the uncertainty in $R$), and the
uncertainty on the $g_2$ correction. Since the knowledge
of $R$ and $g_2$ may improve in the future, we list
the values we used in the final results table. 

\subsubsection{Uncertainty on $D^\prime$}
The dominant uncertainty in $D^\prime$ comes from the
uncertainty in $R$ 
(since $\frac{dD^\prime}{D^\prime}=\epsilon dR$).
Fortunately, relatively recent precision data from JLab 
have reduced the overall uncertainty in $R$ in the
kinematic region of the present experiment 
to approximately 0.03 (see Ref.~\cite{christy}). 
Since $0.4<\epsilon<0.9$
for the present experiment, the uncertainty in $R$
introduces a relative uncertainty of 1\% to 3\% in $g_1/F_1$.
Another source of uncertainty in $D^\prime$ comes from
the estimated systematic uncertainties in the beam energy
of 10 MeV, the electron momentum $P$ of about $0.002P$, and
the scattering angle $\theta$ of about 0.5 mr. Taken
together, these result in a relative uncertainty
in $D^\prime$ of 0.5\% to 1\%, with the largest
uncertainty at large $y$.

\subsubsection{Uncertainty on the $g_2$ correction}
The values of $g_2/g_1$ used in the analysis were
taken from a fit to world data~\cite{nguler}. In
order to estimate the systematic uncertainty on the correction,
we also used the assumption that there are no
deviations from the twist-two model of Wandzura
and  Wilczeck ($g_2^{WW}$)~\cite{WW}.
We estimated the systematic uncertainty on the $g_2$ correction
to be about $\sim$ 3\% for both targets.

\subsection{Physics Results}
The essential physics results from this analysis
are the ratios $g_1/F_1$ for the proton and deuteron,
from which $g_1$ can be obtained by using fits of world data of $F_1$.
These results can be examined as a function of $W$ to look
for resonance structure, and as a function of $x$
to study quark and gluon spin distributions, taking into
account higher-twist contributions.

\subsubsection{$g_1/F_1$ as a function of $W$}
\begin{figure}[hbt]
\centerline{\includegraphics[width= 5in,angle=90] {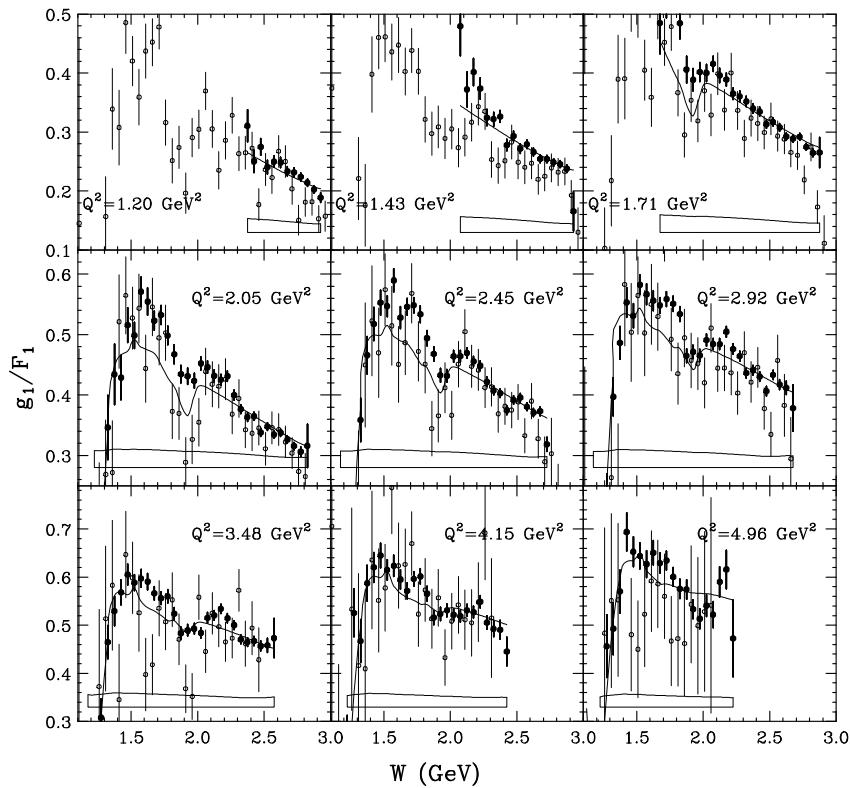}}
\caption{
Results for $g_1/F_1$ as a function of $W$
for the proton in nine bins of $Q^2$. The present
results are shown as solid circles, and the published
Eg1b results are shown as open circles.} 
\label{fig:g1pw}
\end{figure}

The results for $g_1/F_1$ as a function of $W$
are shown for the proton in nine bins of $Q^2$
in Fig.~\ref{fig:g1pw}. 

The systematic uncertainty bands
are dominated by an overall normalization uncertainty
common to all points. The results are in reasonable
agreement with the published results from the
Eg1b experiment~\cite{Eg1b}, but have typically five times
higher statistical precision, and a factor of two better
systematic precision.
The Eg1b results are on average a few percent 
lower than the present results, which is well within the 
overall systematic 
uncertainty of the two experiments (about 3\% for this experiment
and 7\% for Eg1b).

The bands at the bottom of each panel
represent the total systematic uncertainty (point-to-point
as well as overall normalization uncertainties combined). Note the offset from 0
in the vertical axis of most of the panels.

\begin{figure}[hbt]
\centerline{\includegraphics[width=4.5in,angle=90] {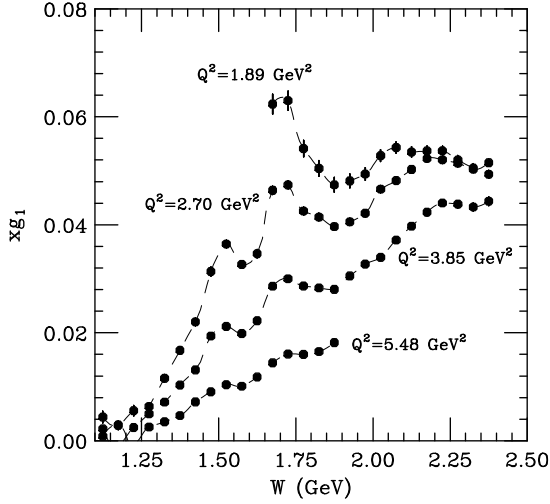}}
\vskip -0.4in
\caption{
Results for $xg_1$ as a function of $W$
for the proton in four bins of $Q^2$. The dashed lines simply
connect the results at a particular value of $Q^2$.
}
\label{fig:g1nof1pw}
\end{figure}

These higher precision results for $g_1/F_1$ 
clearly shows some resonant-like structure for
values of $W$ as high as 2.4 GeV. To see the higher-$W$
structures more clearly, we plot $xg_1$ in
Fig.~\ref{fig:g1nof1pw}. In this plot, pairs of
adjacent $Q^2$ bins have been averaged together.
\vskip -1.2in

The downward trend seen in the $g_1/F_1$ plot, near $W=1.23$ GeV
 can be attributed
to transitions to the spin-3/2 $\Delta(1232)$.
The dip can be understood by realizing that
 $g_1/F_1$ is closely related to 
\be
A_1=\frac{\sigma_{1/2} - \sigma_{3/2}}
         {\sigma_{1/2} + \sigma_{3/2}},
\ee
where $\sigma_{1/2}$ ($\sigma_{3/2}$) 
represent transitions to final state helicity 1/2(3/2).
For a pure $N\rightarrow \Delta$
transition, $A_1=-0.5$.

The strong peak near 1.5 GeV seen in both the $g_1/F_1$ and the $xg_1$ plots is dominated 
by transitions to two nucleon resonances, the $N(1520)$ $3/2^-$ and $N(1535)$ $1/2^-$ states. The 
spin-1/2 $N(1535)$ resonance has only helicity = 1/2 transition amplitudes contributing, 
while the spin-3/2 $N(1520)$ state has contributions from both helicity = 1/2 and 
helicity = 3/2 transition amplitudes, and therefore can exhibit a range of asymmetry values, 
from $A_1 = -1$ for $\sigma_{3/2}$ dominance to $A_1 = +1$ for $\sigma_{1/2}$ dominance.
As was found in the analysis of single pion ($\pi$) and single eta ($\eta$) exclusive cross section 
measurements from other CLAS experiments~\cite{azn1,azn2}, the helicity structure of this state changes 
rapidly from $A_1 = -1$ at $Q^2 = 0$ GeV$^2$ to $A_1 = +1$ at $Q^2>1$ GeV$^2$. In the $Q^2$-range of this 
experiment both states have $A_1 = +1$ to create the strong peak at 1.5 GeV.  

The peak near 1.7 GeV has a similar origin. While more resonances contribute in this 
mass range, the $N(1680)$ $5/2^+$ state was found to be by far the most dominant one for a proton target.
The helicity structure is similar to the $N(1520)$ $3/2^-$ discussed above, and shows a dominant 
helicity 3/2 transition at $Q^2 = 0$ GeV$^2$, which changes to helicity 1/2 dominance for $Q^2>1$ GeV$^2$, 
resulting in asymmetry $A_1 > 0$, growing with $Q^2$.
All of these features have been seen before, at somewhat lower $Q^2$~\cite{Eg1b,RSS}. 
What is  interesting is that all three primary resonance regions are visible for
$Q^2$ values as high as 4 GeV$^2$.
 
What is new in the present results is a confirmation
of an  apparent dip
near $W=1.9$ GeV, with a possible peak near the
$W=2.2$ GeV (sometimes known as the fourth resonance
region). Hints of these structures were seen in
the Eg1b results~\cite{Eg1b}.
The strength of the  dip  near $W = 1.9$ GeV seems to decrease with increasing $Q^2$.
This could indicate significant contributions from resonances with dominant helicity = 3/2 transitions. 
There is a number of states in this mass range that could contribute to the observed structure. 

To investigate this possibility further, we are in the
process of analyzing exclusive single pion ($\pi$) and eta ($\eta$) electroproduction from the
same experiment. The analysis of pion production on neutrons may provide 
information on the isospin of the resonance contribution.

\begin{figure}[hbt]
\centerline{\includegraphics[width=5in,angle=90]{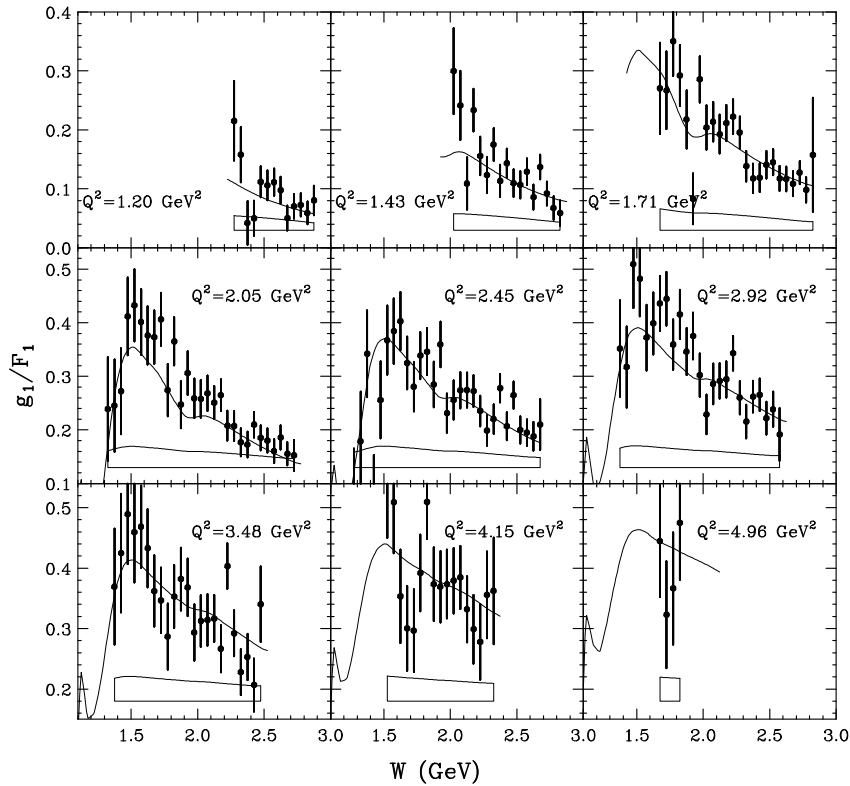}}
\caption{
Same as Fig.~\protect{\ref{fig:g1pw}} except for 
the deuteron. For clarity, data from Eg1b are not shown.
}
\label{fig:g1dw}
\end{figure}

The results for $g_1/F_1$ for the deuteron are shown in 
Fig.~\ref{fig:g1dw}. The comparison to the world
data fit (dominated by Eg1b results~\cite{Eg1b})
is reasonably good. The structure seen in the proton
results is not as pronounced, largely due to the 
factor of six to eight larger error bars, resulting
from 1/3 of the integrated luminosity, a factor of
almost three lower $P_bP_t$, and the NH$_3$ contamination
of the target.

\subsubsection{Results for $g_1/F_1$ as a function of $Q^2$}
\begin{figure}[h]
\centerline{\includegraphics[width=5in,angle=90]{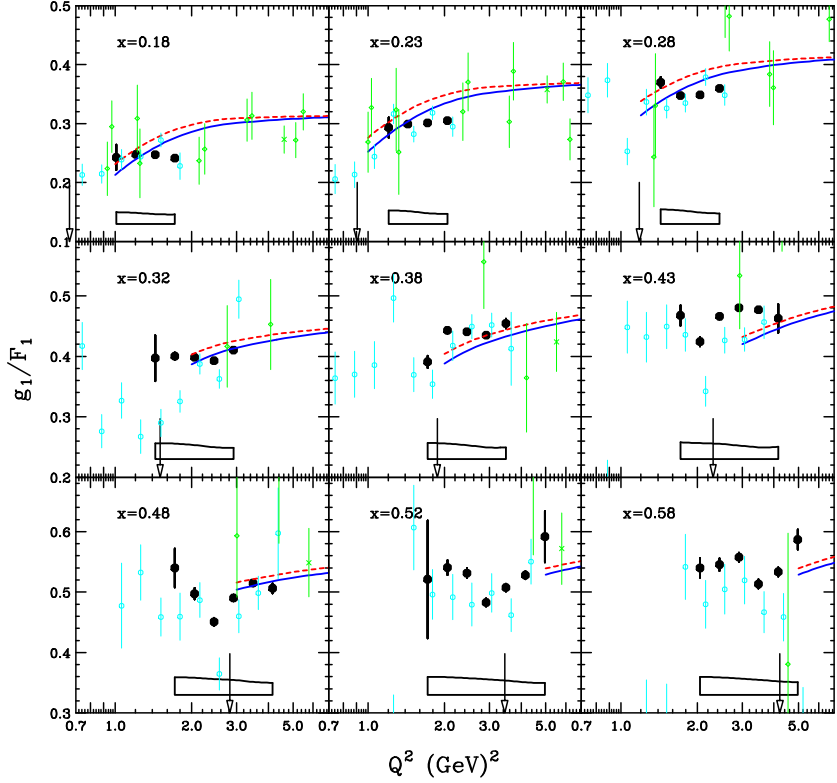}}
\caption{(Color online)
Results for $g_1/F_1$ as a function of $Q^2$
for the proton in nine bins of $x$. 
A bin centering correction has been applied to the data points in this plot, 
of the form $\frac{g_1}{F_1}(x_c,Q^2)/\frac{g_1}{F_1}(x,Q^2)$, where $x_c$ is 
the value of $x$ at the center of the bin, and $x$ and $Q^2$ are the experimental 
averages over the size of the bin. The present
results are shown as the large black solid points, the JLab Eg1b
results~\cite{Eg1b} are shown as the small blue points, and the
results from SLAC~\cite{E143,E155} are shown as the small green points.
The arrows correspond to $W=2$ GeV.}
\label{fig:g1p}
\end{figure}

The results for $g_1/F_1$ as a function of $Q^2$
are shown for the proton in nine bins of $x$
in Fig.~\ref{fig:g1p}. The systematic uncertainty bands
are dominated by an overall normalization uncertainty
common to all points. Small bin-centering corrections
have been applied to the data: typically these are
only significant in the highest and lowest $Q^2$
bins in each panel. 

The bands at the bottom of each panel
represent the total systematic uncertainty (with point-to-point
and overall normalization uncertainties combined).
The dashed red and solid blue curves are representative pQCD
calculation from the LSS group with two models
for gluon polarization ~\cite{LSS}. 

\begin{figure}[h]
\centerline{\includegraphics[width=5in,angle=90]{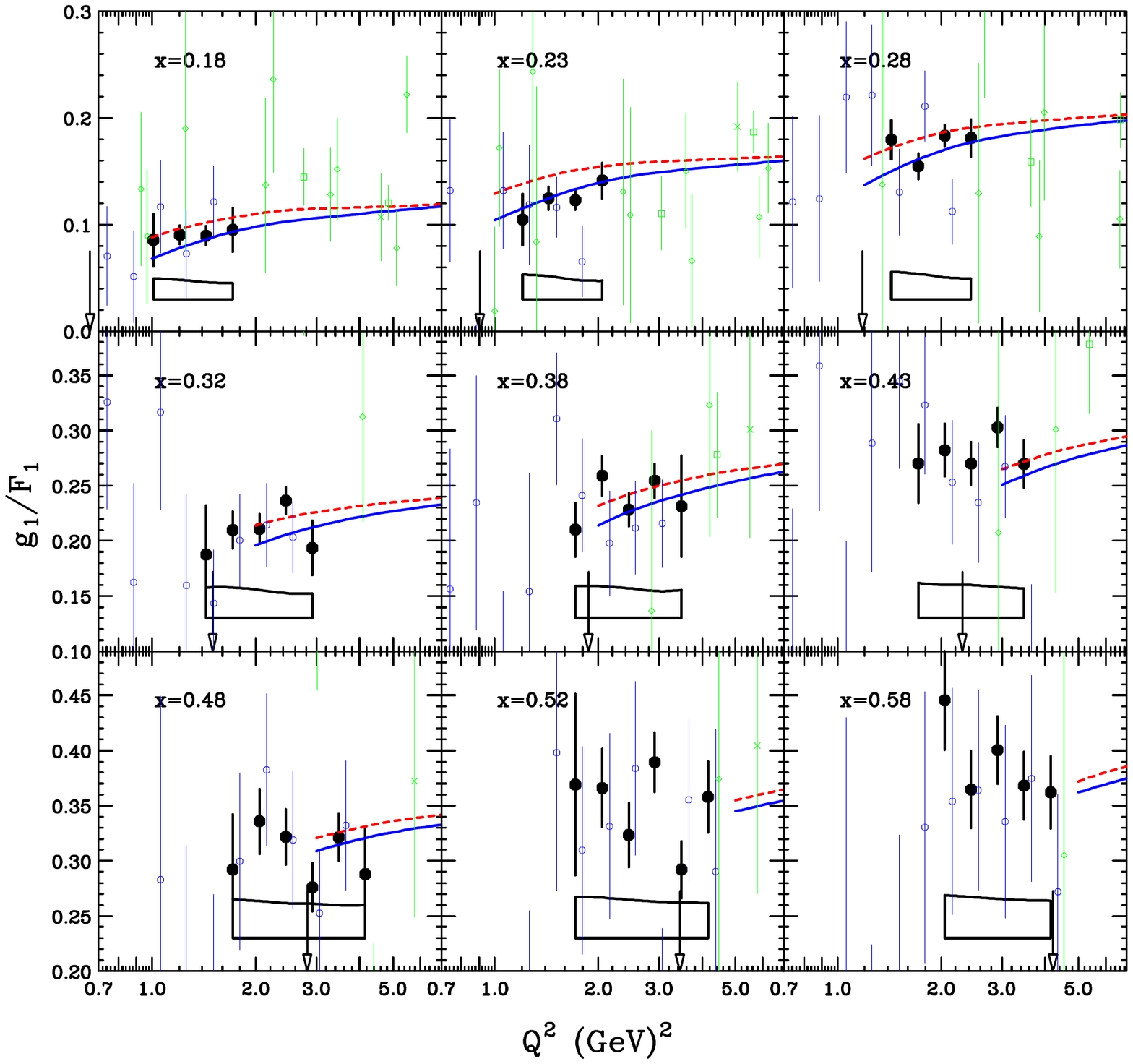}}
\caption{(Color online)
Same as Fig.~\protect{\ref{fig:g1p}} except 
for the deuteron. The small green points include results
from COMPASS~\cite{COMPASS}, HERMES~\cite{HERMES},
and SLAC~\cite{E143,E155}. A bin centering correction has been applied to the data points in this plot, 
of the form $\frac{g_1}{F_1}(x_c,Q^2)/\frac{g_1}{F_1}(x,Q^2)$, where $x_c$ is 
the value of $x$ at the center of the bin, and $x$ and $Q^2$ are the experimental 
averages over the size of the bin.}
\label{fig:g1d}
\end{figure}

The results for the deuteron are shown in 
Fig.~\ref{fig:g1d}. These data are in good
agreement with Eg1b~\cite{Eg1b}
SLAC~\cite{E143,E155}, COMPASS~\cite{COMPASS}, 
and HERMES~\cite{HERMES}. The data are
also reasonably consistent with the fit used
for radiative and other corrections (black curves)
as well as the two pQCD calculations from LSS~\cite{LSS}.
As for the proton case, the deuteron data
show less $Q^2$ dependence than either model, and
are completely consistent with no $Q^2$-dependence
at all for $W>2$ GeV.

The data are consistent with previous
CLAS data with 6 GeV electrons~\cite{Eg1b} (blue points),
and higher beam energy data from SLAC~\cite{E143,E155},
shown as the green points. In the region of overlap, the
present data have much higher statistical and systematic
precision, allowing for a detailed study of the $Q^2$
dependence of $g_1/F_1$ at fixed $x$.

The blue and red curves are pQCD calculations from
the LSS group~\cite{LSS} with positive $\Delta G$ (blue curves) 
and negative $\Delta G$ (red curves). 
In each case, higher twist coefficients 
were fit to give the best agreement with 
the data available in 2007. The 
difference in overall magnitude between 
the curves is of the same order or larger 
than our experimental error bars. 
However, the $Q^2$-dependence is generally larger 
than in the data, and the magnitude of 
the curves is above the data at low $x$ 
and below at higher $x$. It appears that 
the curves with negative $\Delta G$ 
agree better with the flat $Q^2$-dependence 
of our data than the positive $\Delta G$ 
curves. Any such conclusions must be treated
with great caution, because the data clearly
show oscillations around a smooth curve, with a 
dip seen just below the $Q^2$ values corresponding
to $W=2$ GeV, shown as the arrows in the figure panels.


\clearpage

\section{CONCLUSION}\label{s6}

We have made the most precise measurements to date
of $g_1^p$ and $g_1^d$, in
the kinematic range available with a 6 GeV lepton beam.
Examined as a function of $W$, the results show 
resonant structure in the three well-known resonance
regions ($W=1.2$, 1.5, and 1.7 GeV), as well as 
a possible peak near 2.2 GeV, with a dip near 1.9 GeV.
This is the first time some of these resonant structures are
clearly seen at $Q^2$ values well above 1 GeV$^2$.

Examined in the variables $(x,Q^2)$, 
the most striking result is the almost complete lack
of any significant $Q^2$-dependence in the ratio
$g_1/F_1$ for $W>2.2$ GeV and $0.15<x<0.5$. On the
other hand, significant oscillations around a smooth
curve are observed at values of $(x,Q^2)$ corresponding
to $W<2.2$ GeV.

A new global pQCD fit that includes 
our new data should be able to significantly 
improve the determination of higher twist 
corrections, and start to shed light on $\Delta G$.
Such a fit will need to take into account the
effects of nucleon resonance structure below $W$ of
approximately 2.4 GeV. Options include simply
placing a cut $W>2.4$ GeV, using higher-twist
forms that take nucleon resonance structure into
accounts, or smoothing over resonance structure
in some fashion, such as by invoking quark-hadron 
duality in the spin structure functions~\cite{Eg1bduality}.

The availability of an 11 GeV electron beam at
Jefferson Lab, coupled with a new CLAS12 detector,
will soon permit measurements of $g_1$ at higher
values of $W$ with good statistical precision. These
data will be crucial for advancing our knowledge 
of the spin structure of the nucleon.

\section*{Acknowledgments}
We would like to acknowledge the outstanding efforts of the staff
of the Accelerator and the Physics Divisions at Jefferson Lab that made
this experiment possible.  This work was supported in part by
the U.S. Department of Energy and the National
Science Foundation,
the Italian Instituto Nazionale di Fisica Nucleare, the French Centre
National de la Recherche Scientifique, the French Commissariat \`{a}
l'Energie Atomique,the United Kingdom's Science and Technology Facilities Council,
the Chilean CONICYT,   
and the National Research Foundation of Korea.
The Southeastern Universities Research Association (SURA) operates
the Thomas Jefferson National Accelerator Facility for the
United States Department of Energy under contract DE-AC05-84ER-40150.

\newpage
\section{Tables of results}
\begin{table}[hbt]
\begin{tabular}{ccccccccccccc}
x & $Q^2$ & $\epsilon$ & $(g_1/F_1)\pm{\rm stat}\pm{\rm syst}$ & $c_s$ & $f$ & $c_1$ & $P_BP_T$ & $f_{RC}$ & $A_{RC}$  & $D^\prime$ & $\epsilon R$ & $C_{g2}$ \\
 & (GeV)$^2$ & & & & & & & & & \\
\hline
 0.154 &  1.078 &  0.633 & $ 0.222\pm 0.021\pm  0.020$ & 1.013 & 0.150 & -0.005 & 0.641 & 0.919 & -0.013 & 0.686 & 0.171 & 1.007\\
 0.167 &  1.229 &  0.575 & $ 0.240\pm 0.003\pm  0.018$ & 1.018 & 0.152 & -0.006 & 0.641 & 0.919 & -0.016 & 0.737 & 0.148 & 1.009\\
 0.177 &  1.422 &  0.479 & $ 0.249\pm 0.003\pm  0.016$ & 1.042 & 0.154 & -0.007 & 0.641 & 0.903 & -0.023 & 0.812 & 0.117 & 1.014\\
 0.191 &  1.633 &  0.409 & $ 0.257\pm 0.005\pm  0.016$ & 1.067 & 0.156 & -0.007 & 0.641 & 0.886 & -0.032 & 0.865 & 0.094 & 1.000\\
 0.207 &  1.282 &  0.718 & $ 0.278\pm 0.018\pm  0.023$ & 1.005 & 0.152 & -0.007 & 0.642 & 0.950 & -0.008 & 0.628 & 0.171 & 0.998\\
 0.220 &  1.455 &  0.663 & $ 0.295\pm 0.004\pm  0.021$ & 1.006 & 0.154 & -0.008 & 0.641 & 0.950 & -0.010 & 0.686 & 0.150 & 1.000\\
\hline
\end{tabular}
\caption{Results for $g_1/F_1$ for the proton in bins of $(x,Q^2)$, along with average
kinematic values and correction factors for each bin. All values are averaged over the event 
distribution
}
\label{tab:g1px}
\end{table}

\newpage

\begin{table}[hbt]
\begin{tabular}{cccccccccccccc}
x & $Q^2$ & 
$\epsilon$ & 
$(g_1/F_1)\pm{\rm stat}\pm{\rm syst}$ 
& $c_s$ & 
$f$ & 
$c_1$ & 
$P_BP_T$ & 
$f_{RC}$ & 
$A_{RC}$  & 
$D^\prime$ & 
$\epsilon R$ & 
$C_{g2}$ & 
$c_2A_p$ \\
 & (GeV)$^2$ & & & & & & & & & & & \\
\hline
 0.158 &  1.064 &  0.635 & $ 0.072\pm 0.025\pm  0.020$ & 1.012 & 0.213 & -0.044 & 0.225 & 0.943 & -0.014 & 0.686 & 0.170 & 1.022 & 0.020\\
 0.170 &  1.214 &  0.572 & $ 0.087\pm 0.009\pm  0.018$ & 1.021 & 0.214 & -0.044 & 0.224 & 0.941 & -0.016 & 0.740 & 0.146 & 1.024 & 0.024\\
 0.180 &  1.415 &  0.472 & $ 0.093\pm 0.009\pm  0.016$ & 1.046 & 0.216 & -0.044 & 0.223 & 0.926 & -0.021 & 0.819 & 0.115 & 1.030 & 0.028\\
 0.193 &  1.616 &  0.397 & $ 0.108\pm 0.021\pm  0.015$ & 1.078 & 0.217 & -0.044 & 0.223 & 0.909 & -0.029 & 0.874 & 0.091 & 1.000 & 0.031\\
 0.214 &  1.265 &  0.721 & $ 0.097\pm 0.024\pm  0.023$ & 1.004 & 0.212 & -0.044 & 0.224 & 0.962 & -0.011 & 0.626 & 0.173 & 1.006 & 0.024\\
 0.223 &  1.441 &  0.653 & $ 0.123\pm 0.010\pm  0.022$ & 1.007 & 0.214 & -0.044 & 0.224 & 0.961 & -0.013 & 0.696 & 0.148 & 1.008 & 0.028\\
\hline
\end{tabular}
\caption{Results for $g_1/F_1$ for the deuteron in bins of $(x,Q^2)$, 
along with average kinematic values and correction factors for each bin.
All values are averaged over the event distribution.
}
\label{tab:g1dx}
\end{table}

\newpage
\begin{table}[hbt]
\begin{tabular}{ccccccccccccc}
$W$ & $Q^2$ & $\epsilon$ & $(g_1/F_1)\pm{\rm stat}\pm{\rm syst}$ & $c_s$ & $f$ & $c_1$ & $P_BP_T$ & $f_{RC}$ & $A_{RC}$  & $D^\prime$ & $\epsilon R$ & $C_{g2}$ \\
(GeV) & (GeV)$^2$ & & & & & & & & & \\
\hline
 2.987 &  0.634 &  0.413 & $ 0.080\pm 0.092\pm  0.013$ & 1.098 & 0.160 & -0.001 & 0.591 & 0.751 & -0.029 & 0.806 & 0.147 & 1.018\\
 3.019 &  0.610 &  0.388 & $ 0.088\pm 0.058\pm  0.013$ & 1.105 & 0.163 & -0.001 & 0.594 & 0.716 & -0.034 & 0.820 & 0.140 & 1.018\\
 2.888 &  0.752 &  0.480 & $ 0.105\pm 0.083\pm  0.015$ & 1.069 & 0.154 & -0.002 & 0.614 & 0.819 & -0.023 & 0.771 & 0.160 & 1.021\\
 2.930 &  0.738 &  0.446 & $ 0.165\pm 0.033\pm  0.014$ & 1.075 & 0.156 & -0.002 & 0.617 & 0.795 & -0.027 & 0.791 & 0.151 & 1.021\\
 2.979 &  0.727 &  0.403 & $ 0.158\pm 0.023\pm  0.014$ & 1.112 & 0.158 & -0.002 & 0.620 & 0.766 & -0.032 & 0.816 & 0.140 & 1.022\\
 3.007 &  0.705 &  0.382 & $ 0.134\pm 0.042\pm  0.013$ & 1.098 & 0.161 & -0.002 & 0.616 & 0.731 & -0.039 & 0.827 & 0.135 & 1.022\\
\end{tabular}
\end{table}
\newpage
\begin{table}[hbt]
\begin{tabular}{ccccccccccccc}
 2.776 &  1.436 &  0.455 & $ 0.249\pm 0.006\pm  0.016$ & 1.046 & 0.154 & -0.007 & 0.641 & 0.891 & -0.025 & 0.829 & 0.112 & 1.015\\
 2.826 &  1.436 &  0.414 & $ 0.245\pm 0.006\pm  0.015$ & 1.067 & 0.155 & -0.007 & 0.641 & 0.872 & -0.030 & 0.853 & 0.104 & 1.017\\
 2.870 &  1.415 &  0.379 & $ 0.238\pm 0.007\pm  0.015$ & 1.085 & 0.156 & -0.006 & 0.641 & 0.850 & -0.036 & 0.870 & 0.098 & 1.018\\
 2.902 &  1.339 &  0.369 & $ 0.166\pm 0.033\pm  0.014$ & 1.080 & 0.157 & -0.006 & 0.643 & 0.824 & -0.043 & 0.871 & 0.099 & 1.020\\
 1.676 &  1.856 &  0.886 & $ 0.485\pm 0.053\pm  0.030$ & 1.002 & 0.181 & -0.011 & 0.642 & 0.971 & -0.003 & 0.462 & 0.212 & 0.964\\
 1.728 &  1.826 &  0.879 & $ 0.536\pm 0.043\pm  0.029$ & 1.002 & 0.182 & -0.011 & 0.643 & 0.973 & -0.006 & 0.463 & 0.227 & 0.961\\
 1.774 &  1.816 &  0.870 & $ 0.588\pm 0.038\pm  0.028$ & 1.002 & 0.166 & -0.012 & 0.643 & 0.972 & -0.010 & 0.469 & 0.234 & 0.963\\
 1.823 &  1.809 &  0.861 & $ 0.485\pm 0.027\pm  0.028$ & 1.002 & 0.163 & -0.011 & 0.643 & 0.972 & -0.011 & 0.474 & 0.250 & 0.966\\
\end{tabular}
\end{table}
\newpage
\begin{table}[hbt]
\begin{tabular}{ccccccccccccc}
 1.976 &  2.054 &  0.799 & $ 0.423\pm 0.012\pm  0.026$ & 1.004 & 0.167 & -0.010 & 0.641 & 0.977 & -0.002 & 0.565 & 0.204 & 0.968\\
 2.026 &  2.048 &  0.784 & $ 0.452\pm 0.011\pm  0.026$ & 1.005 & 0.165 & -0.010 & 0.641 & 0.976 &  0.000 & 0.590 & 0.181 & 0.971\\
 2.076 &  2.049 &  0.766 & $ 0.446\pm 0.010\pm  0.026$ & 1.004 & 0.163 & -0.010 & 0.641 & 0.975 & -0.001 & 0.616 & 0.161 & 0.974\\
 2.126 &  2.042 &  0.749 & $ 0.432\pm 0.010\pm  0.025$ & 1.004 & 0.162 & -0.010 & 0.641 & 0.974 & -0.002 & 0.639 & 0.147 & 0.976\\
 2.175 &  2.041 &  0.729 & $ 0.425\pm 0.009\pm  0.025$ & 1.003 & 0.161 & -0.010 & 0.641 & 0.973 & -0.003 & 0.661 & 0.137 & 0.979\\
 2.225 &  2.046 &  0.707 & $ 0.431\pm 0.009\pm  0.024$ & 1.002 & 0.160 & -0.010 & 0.641 & 0.971 & -0.004 & 0.684 & 0.129 & 0.981\\
 2.275 &  2.048 &  0.684 & $ 0.400\pm 0.008\pm  0.023$ & 1.003 & 0.160 & -0.010 & 0.641 & 0.970 & -0.005 & 0.705 & 0.123 & 0.984\\
 2.325 &  2.032 &  0.661 & $ 0.377\pm 0.008\pm  0.023$ & 1.003 & 0.159 & -0.009 & 0.641 & 0.968 & -0.007 & 0.724 & 0.119 & 0.986\\
\end{tabular}
\end{table}
\newpage
\begin{table}[hbt]
\begin{tabular}{ccccccccccccc}
 2.475 &  2.427 &  0.518 & $ 0.392\pm 0.007\pm  0.019$ & 1.016 & 0.162 & -0.010 & 0.641 & 0.956 & -0.015 & 0.844 & 0.084 & 0.987\\
 2.525 &  2.421 &  0.485 & $ 0.396\pm 0.007\pm  0.018$ & 1.023 & 0.162 & -0.009 & 0.641 & 0.950 & -0.018 & 0.862 & 0.080 & 0.989\\
 2.574 &  2.425 &  0.449 & $ 0.381\pm 0.007\pm  0.018$ & 1.035 & 0.162 & -0.009 & 0.641 & 0.942 & -0.021 & 0.882 & 0.075 & 0.992\\
 2.625 &  2.430 &  0.410 & $ 0.371\pm 0.007\pm  0.017$ & 1.053 & 0.162 & -0.009 & 0.641 & 0.932 & -0.025 & 0.902 & 0.070 & 0.995\\
 2.673 &  2.411 &  0.375 & $ 0.373\pm 0.007\pm  0.018$ & 1.075 & 0.161 & -0.009 & 0.641 & 0.918 & -0.031 & 0.918 & 0.065 & 0.997\\
 2.715 &  2.317 &  0.357 & $ 0.318\pm 0.012\pm  0.018$ & 1.082 & 0.161 & -0.009 & 0.641 & 0.900 & -0.039 & 0.922 & 0.065 & 1.000\\
 1.175 &  2.903 &  0.884 & $ 0.183\pm 0.055\pm  0.027$ & 1.001 & 0.140 &  0.084 & 0.641 & 0.763 & -0.133 & 0.519 & 0.233 & 0.875\\
 1.276 &  2.906 &  0.872 & $ 0.236\pm 0.034\pm  0.027$ & 1.001 & 0.170 & -0.011 & 0.641 & 0.897 & -0.007 & 0.561 & 0.158 & 0.925\\
\end{tabular}
\end{table}
\newpage
\begin{table}[hbt]
\begin{tabular}{ccccccccccccc}
 1.576 &  3.435 &  0.775 & $ 0.597\pm 0.019\pm  0.028$ & 1.005 & 0.180 & -0.014 & 0.641 & 0.971 & -0.008 & 0.694 & 0.105 & 0.921\\
 1.626 &  3.441 &  0.762 & $ 0.591\pm 0.017\pm  0.027$ & 1.004 & 0.178 & -0.013 & 0.641 & 0.974 & -0.007 & 0.719 & 0.083 & 0.926\\
 1.676 &  3.443 &  0.748 & $ 0.566\pm 0.014\pm  0.027$ & 1.004 & 0.200 & -0.011 & 0.641 & 0.979 & -0.006 & 0.735 & 0.076 & 0.922\\
 1.725 &  3.435 &  0.735 & $ 0.556\pm 0.013\pm  0.026$ & 1.004 & 0.201 & -0.011 & 0.641 & 0.981 & -0.008 & 0.739 & 0.083 & 0.922\\
 1.775 &  3.439 &  0.719 & $ 0.560\pm 0.014\pm  0.026$ & 1.003 & 0.186 & -0.012 & 0.641 & 0.980 & -0.014 & 0.740 & 0.099 & 0.928\\
 1.825 &  3.437 &  0.702 & $ 0.524\pm 0.013\pm  0.025$ & 1.002 & 0.182 & -0.012 & 0.641 & 0.980 & -0.011 & 0.745 & 0.110 & 0.932\\
 1.876 &  3.447 &  0.683 & $ 0.483\pm 0.013\pm  0.025$ & 1.002 & 0.182 & -0.012 & 0.641 & 0.981 & -0.016 & 0.756 & 0.113 & 0.937\\
 1.926 &  3.440 &  0.664 & $ 0.490\pm 0.012\pm  0.024$ & 1.002 & 0.182 & -0.012 & 0.641 & 0.981 & -0.017 & 0.770 & 0.109 & 0.942\\
\end{tabular}
\end{table}
\newpage
\begin{table}[hbt]
\begin{tabular}{ccccccccccccc}
 1.576 &  4.906 &  0.611 & $ 0.627\pm 0.028\pm  0.025$ & 1.004 & 0.182 & -0.013 & 0.641 & 0.971 & -0.014 & 0.886 & 0.059 & 0.910\\
 1.626 &  4.898 &  0.595 & $ 0.651\pm 0.025\pm  0.025$ & 1.002 & 0.182 & -0.013 & 0.641 & 0.973 & -0.013 & 0.909 & 0.041 & 0.914\\
 1.676 &  4.913 &  0.574 & $ 0.629\pm 0.021\pm  0.024$ & 1.002 & 0.203 & -0.011 & 0.641 & 0.978 & -0.014 & 0.927 & 0.033 & 0.913\\
 1.725 &  4.877 &  0.559 & $ 0.635\pm 0.020\pm  0.023$ & 1.003 & 0.206 & -0.011 & 0.641 & 0.979 & -0.011 & 0.930 & 0.036 & 0.913\\
 1.776 &  4.855 &  0.542 & $ 0.600\pm 0.020\pm  0.023$ & 1.004 & 0.193 & -0.012 & 0.642 & 0.978 & -0.019 & 0.930 & 0.043 & 0.917\\
 1.826 &  4.860 &  0.519 & $ 0.576\pm 0.020\pm  0.022$ & 1.007 & 0.191 & -0.012 & 0.642 & 0.977 & -0.019 & 0.937 & 0.047 & 0.921\\
 1.875 &  4.844 &  0.499 & $ 0.574\pm 0.020\pm  0.022$ & 1.008 & 0.191 & -0.012 & 0.642 & 0.977 & -0.021 & 0.943 & 0.048 & 0.925\\
 1.924 &  4.811 &  0.480 & $ 0.533\pm 0.020\pm  0.022$ & 1.011 & 0.191 & -0.012 & 0.642 & 0.975 & -0.020 & 0.951 & 0.047 & 0.929\\
\hline
\end{tabular}
\caption{Results for $g_1/F_1$ for the proton in bins of $(W,Q^2)$, along with average
kinematic values and correction factors for each bin. All values are averages over the event 
distribution.
}
\label{tab:g1pw}
\end{table}

\newpage

\begin{table}[hbt]
\begin{tabular}{cccccccccccccc}
$W$ & $Q^2$ & $\epsilon$ & $(g_1/F_1)\pm{\rm stat}\pm{\rm syst}$ & $c_s$ & 
$f$ & $c_1$ & $P_BP_T$ & $f_{RC}$ & $A_{RC}$  & $D^\prime$ & 
$\epsilon R$ & $C_{g2}$ & $c_2A_p$ \\
(GeV) & (GeV)$^2$ & & & & & & & & & & & \\
\hline
 2.932 &  0.620 &  0.420 & $ 0.034\pm 0.057\pm  0.013$ & 1.072 & 0.220 & -0.044 & 0.236 & 0.825 & -0.023 & 0.804 & 0.146 & 1.164 & 0.000\\
 2.969 &  0.595 &  0.390 & $ 0.100\pm 0.045\pm  0.012$ & 1.086 & 0.223 & -0.044 & 0.236 & 0.797 & -0.026 & 0.821 & 0.138 & 1.166 & 0.004\\
 2.830 &  0.744 &  0.489 & $ 0.132\pm 0.047\pm  0.015$ & 1.051 & 0.216 & -0.044 & 0.233 & 0.876 & -0.020 & 0.768 & 0.159 & 1.087 & 0.000\\
 2.877 &  0.718 &  0.453 & $ 0.028\pm 0.033\pm  0.014$ & 1.059 & 0.217 & -0.044 & 0.233 & 0.858 & -0.022 & 0.789 & 0.151 & 1.090 & 0.005\\
 2.926 &  0.711 &  0.410 & $-0.006\pm 0.028\pm  0.012$ & 1.075 & 0.220 & -0.044 & 0.233 & 0.836 & -0.025 & 0.814 & 0.140 & 1.092 & 0.019\\
 2.957 &  0.694 &  0.383 & $-0.025\pm 0.049\pm  0.012$ & 1.101 & 0.222 & -0.044 & 0.233 & 0.808 & -0.029 & 0.829 & 0.133 & 1.096 & 0.016\\
\end{tabular}
\end{table}
\newpage
\begin{table}[hbt]
\begin{tabular}{cccccccccccccc}
 1.876 &  1.719 &  0.844 & $ 0.218\pm 0.049\pm  0.030$ & 1.004 & 0.214 & -0.044 & 0.223 & 0.976 & -0.015 & 0.487 & 0.264 & 0.977 & 0.032\\
 1.925 &  1.708 &  0.833 & $ 0.083\pm 0.043\pm  0.029$ & 1.004 & 0.214 & -0.044 & 0.223 & 0.976 & -0.015 & 0.501 & 0.257 & 0.982 & 0.031\\
 1.976 &  1.702 &  0.819 & $ 0.286\pm 0.039\pm  0.029$ & 1.005 & 0.213 & -0.044 & 0.223 & 0.976 & -0.009 & 0.524 & 0.235 & 0.982 & 0.031\\
 2.025 &  1.702 &  0.804 & $ 0.204\pm 0.038\pm  0.029$ & 1.005 & 0.213 & -0.044 & 0.223 & 0.976 & -0.004 & 0.550 & 0.209 & 0.983 & 0.031\\
 2.074 &  1.705 &  0.787 & $ 0.214\pm 0.034\pm  0.028$ & 1.004 & 0.213 & -0.044 & 0.223 & 0.976 & -0.005 & 0.576 & 0.188 & 0.984 & 0.034\\
 2.126 &  1.699 &  0.769 & $ 0.193\pm 0.033\pm  0.028$ & 1.004 & 0.214 & -0.044 & 0.223 & 0.975 & -0.006 & 0.600 & 0.171 & 0.986 & 0.033\\
 2.175 &  1.718 &  0.747 & $ 0.212\pm 0.031\pm  0.027$ & 1.003 & 0.213 & -0.044 & 0.224 & 0.974 & -0.008 & 0.627 & 0.158 & 0.989 & 0.034\\
 2.226 &  1.713 &  0.727 & $ 0.223\pm 0.030\pm  0.026$ & 1.002 & 0.213 & -0.044 & 0.224 & 0.973 & -0.009 & 0.649 & 0.149 & 0.992 & 0.033\\
\end{tabular}
\end{table}
\newpage
\begin{table}[hbt]
\begin{tabular}{cccccccccccccc}
 2.525 &  2.039 &  0.510 & $ 0.180\pm 0.023\pm  0.020$ & 1.018 & 0.216 & -0.044 & 0.222 & 0.955 & -0.018 & 0.834 & 0.095 & 1.002 & 0.041\\
 2.575 &  2.038 &  0.474 & $ 0.160\pm 0.023\pm  0.019$ & 1.031 & 0.216 & -0.044 & 0.222 & 0.949 & -0.021 & 0.855 & 0.089 & 1.005 & 0.040\\
 2.624 &  2.030 &  0.438 & $ 0.186\pm 0.023\pm  0.018$ & 1.048 & 0.216 & -0.044 & 0.223 & 0.940 & -0.024 & 0.874 & 0.084 & 1.008 & 0.041\\
 2.674 &  2.032 &  0.397 & $ 0.156\pm 0.021\pm  0.017$ & 1.070 & 0.216 & -0.044 & 0.223 & 0.929 & -0.027 & 0.895 & 0.077 & 1.011 & 0.041\\
 2.720 &  1.977 &  0.368 & $ 0.153\pm 0.029\pm  0.016$ & 1.094 & 0.217 & -0.044 & 0.223 & 0.915 & -0.031 & 0.906 & 0.074 & 1.014 & 0.040\\
 1.278 &  2.460 &  0.893 & $ 0.067\pm 0.099\pm  0.028$ & 1.001 & 0.230 & -0.044 & 0.223 & 0.910 & -0.009 & 0.531 & 0.146 & 0.931 & 0.015\\
 1.324 &  2.410 &  0.891 & $ 0.179\pm 0.093\pm  0.031$ & 1.002 & 0.222 & -0.044 & 0.223 & 0.926 &  0.009 & 0.541 & 0.120 & 0.935 & 0.024\\
 1.375 &  2.432 &  0.883 & $ 0.342\pm 0.081\pm  0.035$ & 1.002 & 0.216 & -0.044 & 0.223 & 0.938 &  0.014 & 0.552 & 0.116 & 0.937 & 0.030\\
\end{tabular}
\end{table}
\newpage
\begin{table}[hbt]
\begin{tabular}{cccccccccccccc}
 1.877 &  2.900 &  0.725 & $ 0.346\pm 0.044\pm  0.033$ & 1.002 & 0.222 & -0.044 & 0.222 & 0.980 & -0.014 & 0.691 & 0.149 & 0.950 & 0.053\\
 1.925 &  2.893 &  0.709 & $ 0.375\pm 0.043\pm  0.032$ & 1.002 & 0.220 & -0.044 & 0.222 & 0.980 & -0.016 & 0.706 & 0.144 & 0.953 & 0.056\\
 1.975 &  2.898 &  0.689 & $ 0.302\pm 0.041\pm  0.032$ & 1.001 & 0.220 & -0.044 & 0.222 & 0.980 & -0.009 & 0.729 & 0.129 & 0.956 & 0.055\\
 2.025 &  2.876 &  0.670 & $ 0.229\pm 0.037\pm  0.031$ & 1.002 & 0.221 & -0.044 & 0.222 & 0.980 & -0.006 & 0.750 & 0.115 & 0.959 & 0.058\\
 2.074 &  2.871 &  0.649 & $ 0.286\pm 0.038\pm  0.031$ & 1.002 & 0.220 & -0.044 & 0.222 & 0.979 & -0.008 & 0.772 & 0.103 & 0.961 & 0.058\\
 2.124 &  2.874 &  0.625 & $ 0.291\pm 0.033\pm  0.030$ & 1.003 & 0.220 & -0.044 & 0.222 & 0.977 & -0.010 & 0.795 & 0.093 & 0.964 & 0.060\\
 2.176 &  2.895 &  0.596 & $ 0.295\pm 0.034\pm  0.029$ & 1.005 & 0.220 & -0.044 & 0.222 & 0.976 & -0.011 & 0.819 & 0.084 & 0.967 & 0.063\\
 2.225 &  2.889 &  0.570 & $ 0.343\pm 0.032\pm  0.028$ & 1.007 & 0.221 & -0.044 & 0.222 & 0.974 & -0.013 & 0.837 & 0.079 & 0.970 & 0.060\\
\end{tabular}
\end{table}
\newpage
\begin{table}[hbt]
\begin{tabular}{cccccccccccccc}
 1.826 &  4.107 &  0.594 & $ 0.509\pm 0.061\pm  0.036$ & 1.003 & 0.229 & -0.044 & 0.222 & 0.979 & -0.011 & 0.868 & 0.071 & 0.931 & 0.082\\
 1.877 &  4.111 &  0.571 & $ 0.373\pm 0.061\pm  0.035$ & 1.004 & 0.228 & -0.044 & 0.222 & 0.979 & -0.016 & 0.878 & 0.073 & 0.934 & 0.079\\
 1.925 &  4.132 &  0.545 & $ 0.369\pm 0.059\pm  0.035$ & 1.006 & 0.227 & -0.044 & 0.222 & 0.978 & -0.016 & 0.894 & 0.069 & 0.937 & 0.082\\
 1.975 &  4.094 &  0.526 & $ 0.374\pm 0.057\pm  0.034$ & 1.009 & 0.225 & -0.044 & 0.221 & 0.976 & -0.015 & 0.906 & 0.064 & 0.941 & 0.082\\
 2.025 &  4.086 &  0.501 & $ 0.379\pm 0.054\pm  0.033$ & 1.012 & 0.225 & -0.044 & 0.220 & 0.974 & -0.014 & 0.922 & 0.057 & 0.944 & 0.082\\
 2.076 &  4.069 &  0.475 & $ 0.385\pm 0.052\pm  0.033$ & 1.017 & 0.225 & -0.044 & 0.220 & 0.972 & -0.017 & 0.937 & 0.051 & 0.947 & 0.081\\
 2.126 &  4.083 &  0.444 & $ 0.332\pm 0.055\pm  0.032$ & 1.022 & 0.227 & -0.044 & 0.218 & 0.969 & -0.021 & 0.954 & 0.046 & 0.950 & 0.084\\
 2.175 &  4.053 &  0.418 & $ 0.299\pm 0.057\pm  0.031$ & 1.029 & 0.226 & -0.044 & 0.217 & 0.965 & -0.024 & 0.965 & 0.043 & 0.953 & 0.083\\
\hline
\end{tabular}
\caption{Results for $g_1/F_1$ for the deuteron in bins of $(W,Q^2)$, along with average
kinematic values and correction factors for each bin. All values are averaged over the event 
distribution.
}
\label{tab:g1dw}
\end{table}

\end{document}